\documentclass[fleqn,usenatbib]{mnras}  
\usepackage{fix-cm}
\usepackage{graphicx}
\usepackage{dcolumn}
\usepackage{bm}
\usepackage{lipsum}
\usepackage{ragged2e}
\usepackage{graphicx}
\usepackage{booktabs}
\usepackage{makecell}
\usepackage{multirow} 
\usepackage{makecell}
\usepackage{makecell}
\usepackage{gensymb}
\usepackage{mathptmx}

\usepackage[T1]{fontenc}

\usepackage{newtxtext,newtxmath}
\usepackage{mathptmx}
\usepackage[T1]{fontenc}
\usepackage{lineno}

\usepackage{orcidlink}

\usepackage{graphicx}	
\usepackage{amsmath}	
\usepackage{hyperref}
\usepackage{comment}

\title{Magnetic field Topology and Star Formation in the Cepheus B Filamentary cloud under External Feedback}

\author[Sandhyarani et al.]{
Panigrahy Sandhyarani$^{1}$\thanks{E-mail: sandhyaranipanigrahy@students.iisertirupati.ac.in}\orcidlink{0009-0007-6357-6874}, 
Chakali Eswaraiah$^{2,1}$\thanks{Corresponding author e-mail: eswaraiahc@iisermohali.ac.in}\orcidlink{0000-0003-4761-6139}, 
Manash R. Samal$^{3}$\orcidlink{0000-0002-9431-6297}, 
Vineet Rawat$^{3,4}$\orcidlink{0000-0002-6828-5108}, 
Jihye Hwang$^{5,6}$\orcidlink{0000-0001-7866-2686}, 
\newauthor
Jia-Wei Wang$^{7}$\orcidlink{0000-0002-6668-974X}, 
Yuehui Ma$^{8}$\orcidlink{0000-0002-8051-5228}, 
Patricio Sanhueza$^{9}$\orcidlink{0000-0002-7125-7685}, 
Jessy Jose$^{1}$\orcidlink{0000-0003-4908-4404}
\\
$^{1}$Department of Physics, Indian Institute of Science Education and Research Tirupati, Yerpedu, Tirupati 517619, Andhra Pradesh, India\\
$^{2}$Department of Physical Sciences, Indian Institute of Science Education and Research (IISER) Mohali, Knowledge City, Sector 81, SAS Nagar 140306, Punjab, India\\
$^{3}$Physical Research Laboratory (PRL), Navrangpura, Ahmedabad 380009, Gujarat, India\\
$^{4}$ Korea Astronomy and Space Science Institute (KASI), Yuseong-gu, Daejeon 34055, Republic of Korea\\
$^{5}$Institute for Advanced Study, Kyushu University, Japan\\
$^{6}$Department of Earth and Planetary Sciences, Faculty of Science, Kyushu University, Nishi-ku, Fukuoka 819-0395, Japan\\
$^{7}$East Asian Observatory, 660 N. A\'{o}h\={o}k\={u} Place, University Park, Hilo, HI 96720, USA\\
$^{8}$Purple Mountain Observatory and Key Laboratory of Radio Astronomy, Chinese Academy of Sciences, 10 Yuanhua Road, Nanjing 210033, People's Republic of China\\
$^{9}$Department of Astronomy, School of Science, The University of Tokyo, 7-3-1 Hongo, Bunkyo, Tokyo 113-0033, Japan
}

\date{Accepted XXX. Received YYY; in original form ZZZ}

\pubyear{2025}

\begin{document}
\label{firstpage}
\pagerange{\pageref{firstpage}--\pageref{lastpage}}
\maketitle

\begin{abstract}
We present a detailed study of the Cep B molecular cloud based on sub-mm dust polarization and $^{13}$CO (J=3–2) spectral line observations obtained with SCUBA-2/POL-2 and HARP on the James Clerk Maxwell Telescope (JCMT). The 850 $\mu$m dust continuum map reveals a prominent filamentary structure oriented Northwest–Southeast (NW–SE), with the magnetic field (B-field) displaying a distinct morphology~--~curving into a bow-like shape near the filament head and aligning along the spine toward the tail. The filament is thermally supercritical, with its line mass exceeding the critical value for an isothermal filament, indicating that self-gravity drives radial contraction. The mass-to-flux ratio suggests that the filament is magnetically subcritical on global scales, implying that B-fields provide significant support against collapse. Despite this, the presence of dense cores and embedded star formation indicates that collapse proceeds locally. The observed core spacing spans a range of values, with the largest separations comparable to the expected fragmentation scale for a self-gravitating filament undergoing sausage instability, suggesting that gravitational instability sets the primary fragmentation scale. Smaller separations and non-uniform spacing may indicate the influence of local variations and hierarchical fragmentation.  Overall, Cep B represents a system in which gravity drives fragmentation, B-fields regulate its evolution, and external feedback shapes both its morphology and star formation activity at the head of the filament.

\end{abstract}




\section{Introduction}

Stars predominantly form within filamentary structures in molecular clouds, as revealed by far-infrared and submillimeter observations~\citep{andre2010,Arzoumanian2011}. These filaments serve as the primary sites where dense cores emerge and subsequently evolve into stars~\citep{andre2014,konyves2015}. Understanding how filaments form, evolve, and fragment is therefore central to building a comprehensive picture of star formation. In idealized conditions, self-gravity can lead to the fragmentation of filaments into regularly spaced dense cores through gravitational instabilities, often referred to as the “sausage” instability~\citep{chandrasekhar1953problems,nagasawa1987gravitational}. However, observations show that real filaments often exhibit complex and non-uniform fragmentation patterns, indicating that additional physical processes play a crucial role in regulating their evolution~\citep{hacar2013cores,tafalla2015chains}.

Magnetic fields (B-fields) are widely observed in star-forming regions and are often organized with respect to filamentary structures~\citep{planck2016,hull2019interferometric}. Their presence can influence both the morphology and dynamics of molecular clouds, regulating the growth of instabilities and guiding gas flows~\citep{crutcher2012,pattlefissel2019}. The role of B-fields is particularly important in determining how and where fragmentation proceeds, as well as in shaping the resulting distribution of dense cores~\citep{mouschovias1991,henebele2019}. In magnetized environments, the coupling between gas and B-fields can delay or regulate gravitational collapse, while processes such as ambipolar diffusion allow neutrals to gradually decouple from the B-field, enabling localized collapse and star formation~\citep{mouschovias1977connection,mouschovias1979ambipolar,shu1987,mouschovias1992ambipolar,lada2012physics}.

In addition to B-fields, external feedback from massive stars can significantly alter the structure and evolution of molecular clouds. Radiation-driven processes, such as ionization fronts and photoevaporative flows, can compress pre-existing dense material and reshape clouds into elongated or cometary structures~\citep{bertoldi1989photoevaporation,lefloch1994cometary,miao2006triggered,bisbas2011radiation}. Such interactions may trigger or enhance star formation and can modify both the physical conditions and fragmentation properties of filaments~\citep{deharveng2015bipolar,dale2015,samal2018bipolar}.

Cepheus B (Cep B), situated at a distance of approximately 725 parsecs \citep{sargent1977molecular,mookerjea2012structure}, this molecular core has garnered significant attention as a prototype for sequential star formation \citep{testi1995sequential,getman2006chandra}. Its proximity to the Cepheus OB3 association, particularly its younger subgroup Cep OB3b, places it in a dynamic environment where the influence of massive stars can profoundly impact the birth of a secondary generation of stars \citep{blaauw1964associations,kun2008star}.

In this work, we present submillimeter dust polarization and molecular line observations of Cep B obtained with JCMT/SCUBA-2/POL-2 and HARP. We investigate the B-field morphology, the physical properties of the filament, and the nature of its fragmentation. By combining these analyses, we aim to understand how B-fields and external feedback regulate the evolution and star formation activity in the Cep B molecular cloud.

\section{Observations and Data Reduction}\label{data reduction}
\subsection{JCMT SCUBA-2/POL-2 Data}

We carried out polarization continuum observations towards massive star-forming region Cep B with the reference position at $\alpha_{J2000.0}$=$22^h 57^m 07^s$, $\delta_{J2000.0}$=$62 ^{\circ} 37\arcmin 22.3 \arcsec$ with SCUBA-2 POL-2 mounted on the JCMT (project code: M24AP055; PI: Sandhyarani Panigrahy). We have allotted a total observation time of 9h 42m. All these observations were taken in Band-2 weather conditions with opacity $\tau_\mathrm{225GHZ}$ ranging from 0.05 to 0.08.  In this study, each MSB involves a 40-minute observing session using the POL-2 instrument with the POL-2 DAISY scan pattern \citep{friberg2016pol}. A total of 14 MSBs were observed over four days: 29 August, 2 September, 4 September, and 12 September 2023. The obtained data were processed through a two-step reduction procedure utilizing the \textit{pol2map}\footnote{\url{http://starlink.eao.hawaii.edu/docs/sun258.htx/sun258ss73.html}} routine, which was integrated into the Sub-Millimeter User Reduction
Facility (SMURF\footnote{\url{http://starlink.eao.hawaii.edu/docs/sun258.htx/sun258.html}}) package of Starlink software \citep{jenness2015}. Continuum polarization at both 450 and 850 $\mu$m was acquired simultaneously. The 850 $\mu$m data are considered in this paper.
\vskip 2pt
In the initial phase, the raw bolometer timestreams of each observation was transformed into separate timestreams for Stokes $Q$, $U$, and $I$ polarization components. Subsequently, an initial Stokes I map was generated from the I timestream of each observation using the iterative map-making routine called \textit{makemap} \citep{chapin2013scuba}. During the reduction process, regions of astrophysical emission were identified through an iterative signal-to-noise-based masking technique employed by \textit{makemap}. Regions outside this masked area were assigned zero values until the final iteration of the map-making process. The resulting maps were then compared to the first map in the sequence to determine a set of relative pointing corrections. Finally, the individual Stokes I maps were co-added to produce an initial I map representing the region of interest. A comprehensive discussion on the role of masking in SCUBA-2 data reduction is described by \citet{mairs2016jcmt}.

In the second stage of the data processing, an enhanced Stokes I map was generated using the I timestreams from each observation, employing the \textit{makemap} routine. Additionally, Stokes Q and U maps were created from their respective timestreams. The initial I map obtained in the first stage was utilized to establish a fixed signal-to-noise-based mask that remained consistent throughout all iterations of the map-making process. During this second stage, the \textit{skyloop}\footnote{\url{http://starlink.eao.hawaii.edu/docs/sun258.htx/sun258ss72.html}} routine was employed, wherein each observation in the set underwent one iteration of \textit{makemap} consecutively, and the set was then averaged at the conclusion of each iteration. The pointing corrections determined in the first stage were applied to the Stokes Q, U, and I maps during the map-making process. Moreover, the instrumental polarization in the Stokes Q and U maps was corrected based on the final output I map, using the `August 2019' instrumental polarization (IP) model\footnote{\url{ https://www.eaobservatory.org/jcmt/2019/08/new-ip-models-for-pol2-data/}}.
The map-making uncertainty was reduced by using the \textit{skyloop} mode, and the \textit{MAPVARS} mode was used to calculate the overall uncertainty from the standard deviation of the individual observations. A number of B-fields In STar-forming Region Observations (BISTRO) studies, such as \cite{ward2017first,kwon2018first,wang2019jcmt,wang2024filamentary}, provide a detailed description of the steps and procedure involved in the data reduction process. 
\vskip 2pt
Finally obtained Stokes I,
Q, and U maps are gridded to a pixel size of 4$\arcsec$ in units of pW. A pixel size of 4$\arcsec$ was used for POL-2 data reduction, as bigger pixel size can increase uncertainty owing to the map-making process. We converted pW to Jy beam$^{-1}$ by multiplying
the flux conversion factor of 495 Jy beam$^{-1}$ pW$^{-1}$ at SCUBA-2 at 850$\mu$m, which is recently updated by \cite{mairs2021decade}. The POL-2 data are further adjusted by applying an additional correction factor of 1.35, accounting for the reduced optical throughput caused by the insertion of the rotating POL-2 half-wave plate into the SCUBA-2 light path. This results in a final flux conversion factor of 668 Jy beam$^{-1}$ pW$^{-1}$ \citep{mairs2021decade}.  
The unpredictability in POL-2 images results from a combination of instrumental noise and map-making uncertainty. Small perturbations in the input data can be amplified by the nonlinear mapping process, leading to significant differences in the resulting intensity maps. This is considered
as a further source of uncertainty.  The rms noise in the Stokes I image is found to be 3.4 mJy/beam. The calculated polarization fraction P was debiased with the asymptotic estimator \citep{wardle1974linear}

\begin{align}
     \hspace{6em}P &= \frac{1}{I} {\sqrt{{(Q^{2}} + {U^{2}}) - \frac{1}{2} (\sigma_{Q}^{2}+\sigma_{U}^{2})}},
\end{align}


where P is the debiased polarization percentage and I, Q, U, $\sigma_I$, $\sigma_Q$, and $\sigma_U$ are the Stokes I, Q, U, and their uncertainties. The
uncertainty of the polarization fraction was estimated using

\begin{align} 
    \hspace{6em}\sigma_p = \sqrt{\frac{({Q^2} \sigma_Q^2 + U^2 \sigma_U^2)}{I^2(Q^2+U^2)} + \frac{\sigma_I^2 (Q^2 + U^2)} {I^4}} \quad.
\end{align}

The polarization position angles ($\theta$), measured from north toward east in the plane of the sky, were determined using the following relation,
\begin{align*}
    \hspace{6em}\theta = \frac{1}{2} \tan^{-1} \frac{U}{Q}.
\end{align*}

The corresponding uncertainties in $\theta$ were calculated using
\begin{align*}
    \hspace{6em}\sigma_{\theta} = \frac{1}{2} \frac{\sqrt{Q^2\sigma U^2 + U^2\sigma Q^2}}{(Q^2 + U^2)} \times \frac{180\degree}{\pi}.
\end{align*}
\begin{figure*}
\includegraphics[width=0.8\textwidth]{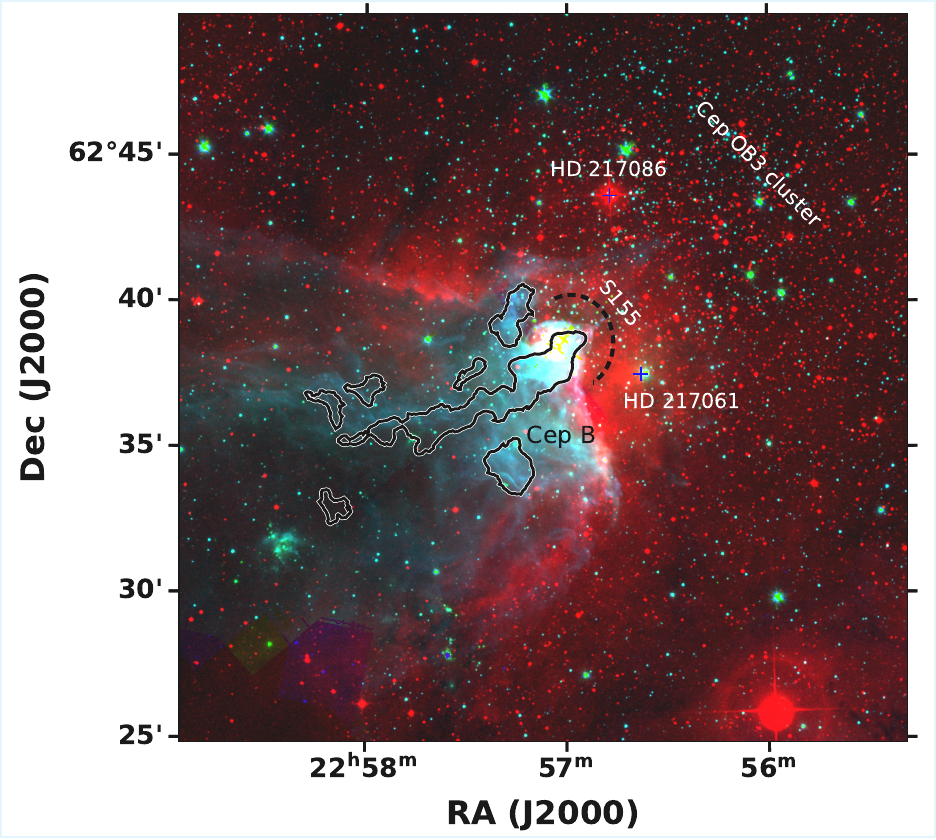}
\caption{The RGB three-colour composite image of Spitzer 3.6 $\mu$m (blue), 4.5 $\mu$m (green), and DSS2-red (red) data showing the Cep OB3 association and the Cep B cloud. The blue `+' sign represents the OB-type stars impacting Cep B and the surrounding regions. The black contour is the same as yellow contour in Figure\ref{fig:Bfieldmorphology_unscaled}.} 
\label{fig:bigpicture}
\end{figure*}
\subsection{HARP molecular line observations}
To understand the gas kinematics within the Cep B region it is necessary to analyze the molecular line observations. These observations offer valuable information on the velocity distribution and dynamic behavior of the molecular gas, thereby enhancing our understanding of the physical processes driving star formation in this region. The $^{13}$CO (J = 3-2) and C$^{18}$O (J = 3-2) observations towards Cep B were conducted simultaneously using the Heterodyne Array Receiver Program (HARP) on the JCMT, under project code M24BP056 (PI: Sandhyarani Panigrahy). The MSBs were observed in 5 days 11th July, 17th July, 18th July, 19th July and 10th August 2024. These spectral lines were observed at a rest frequency of 330.58 GHz. The observations were carried out under Band-2 weather conditions, with the zenith opacity ($\tau$) ranging from 0.05 to 0.08. The mapped area covered a 7\arcmin $\times$ 7\arcmin region, using the Position Switching mode and Boustrophedon scan pattern, with a 250 MHz bandwidth. Data reduction was performed using the ORAC data reduction (ORAC-DR) pipeline and the Kernel Application Package (KAPPA) \citep{currie2008starlink} of the Starlink software \citep{jenness2015}. The pixel size of the reduced C$^{18}$O map is 7.27\arcsec.

\subsection{\textit{Planck} dust polarization data}

The large-scale B-field in the Cep A region is obtained by \textit{Planck} 353 GHz (850$\mu$m) dust continuum polarization data \citep{2016A&A...586A.138P}. The polarization data, including the Stokes I, Q, and U maps, have been extracted from the \textit{Planck} public data Release 2 (PR2 Full Mission map with PCCS2 catalog \footnote{\url{https://irsa.ipac.caltech.edu/applications/planck/}}) \citep{2016A&A...586A.138P}. 
These maps have a beam size of $\sim 5\arcmin$ and a pixel size of $\sim 1\arcmin$. The data have been reduced using the standard
procedures described by \cite{ade2015planck,ade2016planck,soler2016magnetic,baug2020alma} and references therein.

\section{RESULTS} \label{results}
\subsection{JCMT/POL-2 Dust continuum map}
\begin{figure*}
\includegraphics[width=0.9\textwidth]{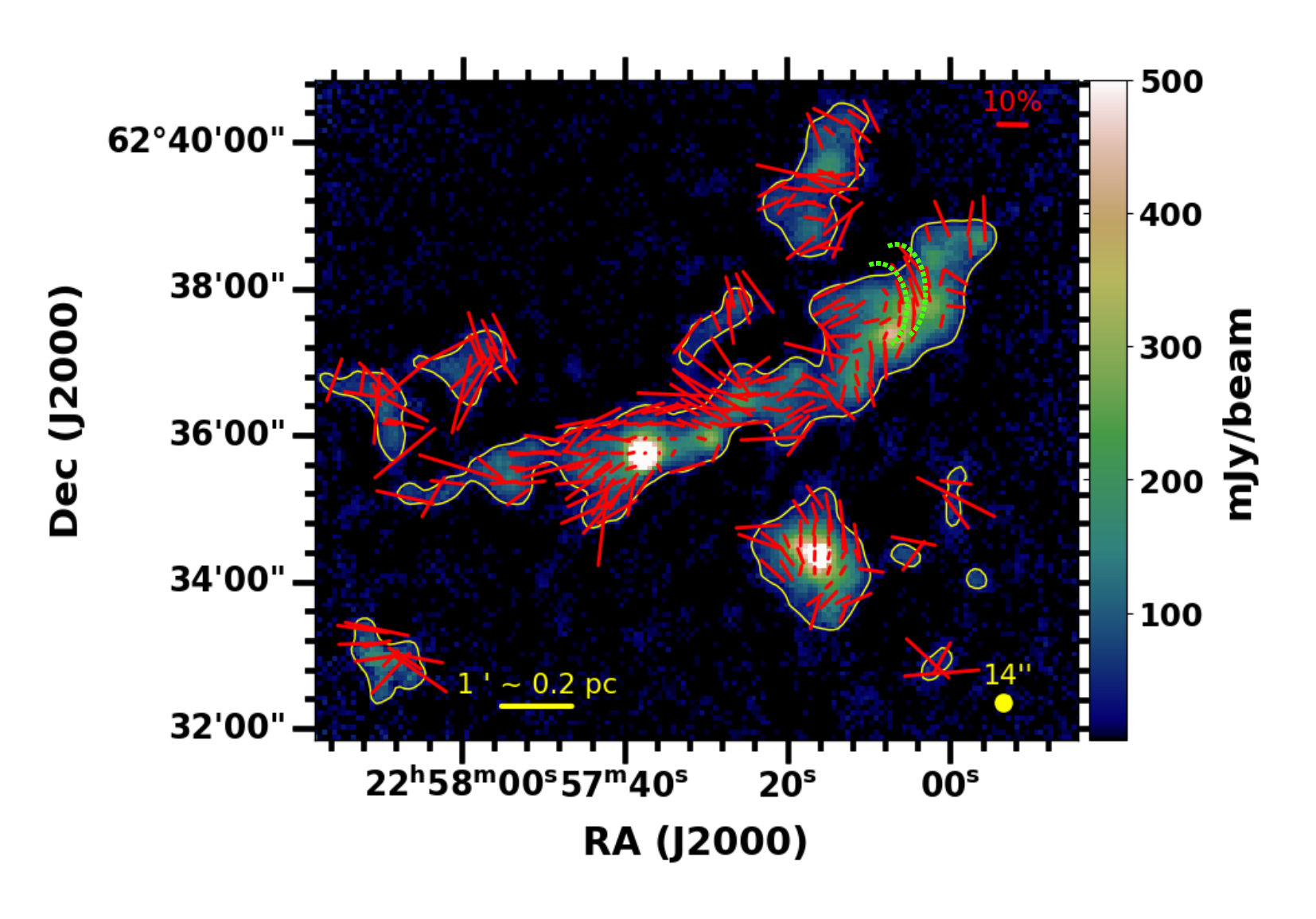}
\caption{JCMT SCUBA-2/POL-2 850 $\mu$m dust continuum map of the Cep A clump with B-field segments overlaid after applying an offset of 90 $^\circ$ to polarization angles. The length of each vector is proportional to the fraction of polarization. Yellow contour is drawn at 35 mJy/beam corresponds to the 10$\sigma$, where $\sigma$ is the rms noise in the dust continuum map and is 3.5 mJy/beam. The green curves shows the bow-like morphology of B-field.} 
\label{fig:Bfieldmorphology_unscaled}
\end{figure*}

Figure \ref{fig:Bfieldmorphology_unscaled} provides a detailed view of the dust continuum map of the massive star-forming clump Cep B at 850 $\mu$m. The map prominently displays an elongated filamentary structure oriented in the Northwest-Southeast (NW-SE) direction, with the cores appearing as brighter regions, fenced by dimmer, more extended structures. A head-like structure is evident at the tip of the NW portion of the region. Overall, Cep B exhibits a head and tail morphology analogous to the cometary globule as shown in Figure \ref{fig:Bfieldmorphology_unscaled} (also see Figure \ref{fig:bigpicture}). Towards the bottom, we can identify one more core with the maximum intensity of 861 mJy/beam. There exist few more structures around the filament. To achieve significant detections, we limit the polarization segments by applying the vector selection criteria I/$\sigma_\mathrm{I}$ $\geq$ 10, PI/$\sigma_\mathrm{PI}$ $\geq$ 2 and P$\leq$ 30\%, we obtained 289 polarization segments. The polarization map at 850 $\mu$m is superimposed on the Stokes I image as shown in Figure \ref{fig:Bfieldmorphology_unscaled}, which has a pixel size of 4$\arcsec$ and a beam size of 14$\arcsec$. The level of noise, or RMS ($\sigma$), in the Stokes I map is approximately 3.4 mJy/beam in the signal-free region. The yellow contours indicate at 34 mJy/beam corresponds to 10$\sigma$ intensity of the 850 $\mu$m continuum emission.

\subsection{B-field morphology}
The observed B-field morphology is shown in Figure \ref{fig:Bfieldmorphology_unscaled}, where each polarization segment has been rotated by 90$^\circ$. The length of each segment represents the polarization fraction, providing insight into the change in the amount of polarization across the region.

\begin{figure*}
\includegraphics[width=0.9\textwidth]{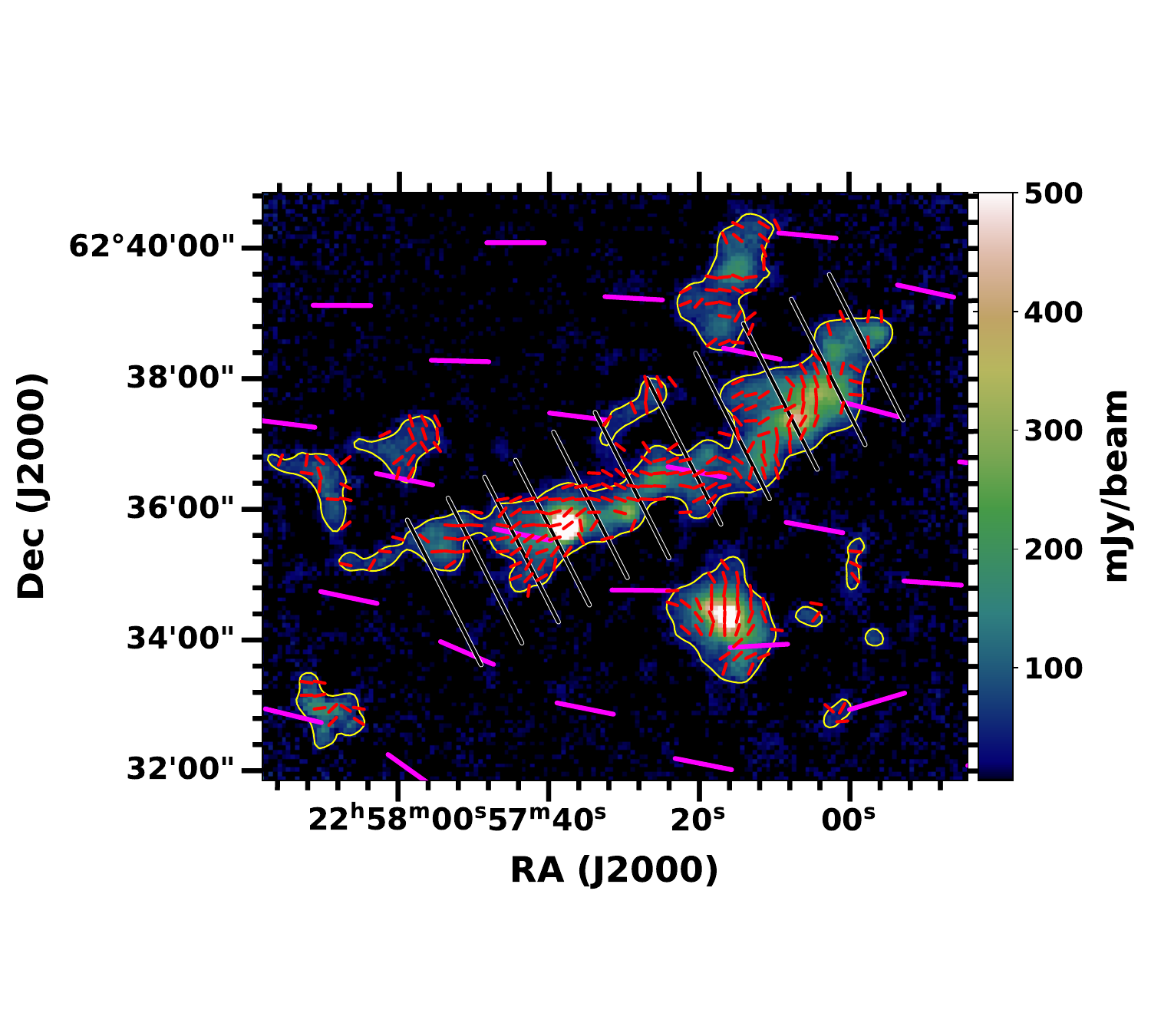}
\caption{Same as Figure \ref{fig:Bfieldmorphology_unscaled} but using equal length for all the B-field segments to better view the B-field morphology. Here magenta segments represent the {\it Planck} B-field. The black lines are the cuts along which the column density profiles are extracted.} 
\label{fig:Bfieldmorphology}
\end{figure*}

\begin{figure}
    \centering
    \includegraphics[width=1\linewidth]{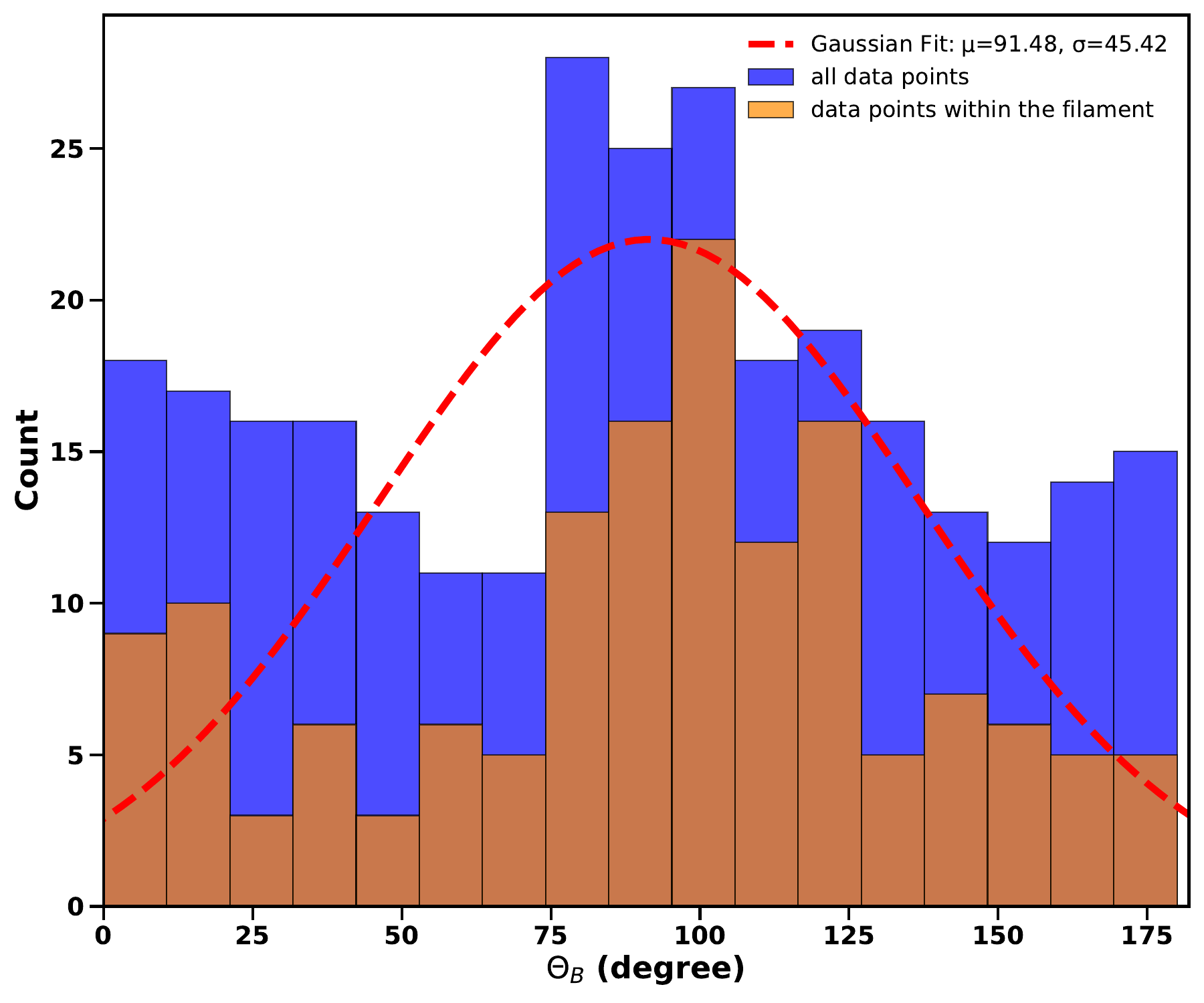}
    \caption{Represents the histogram of B-field position angle (PA) for all the data points (black) and the data points within the filamentary structure (red). The red dashed line represents the gaussian fitted to the PA within the filamentary structure.}
    \label{fig:Gaussianfit}
\end{figure}

Additionally, Figure \ref{fig:Bfieldmorphology} presents an equal-length B-field vector map, with the {\it Planck} B-field vectors overlaid in magenta segments for comparison. From this, we can discern that the POL-2 B-field predominantly aligns with the elongated filamentary structure.
We examine the distribution of B-field position angles (PAs) in Figure \ref{fig:Gaussianfit}, which includes histograms for all the B-field PAs as well as for those B-field PAs confined within the filament. A striking pattern emerges: towards the head of the filamentary cloud, the POL-2 B-field closely follows the orientation of the {\it Planck} B-field, indicating a consistent B-field across different scales. However, as we move along the filament toward the tail, the B-field undergoes a noticeable twist, deviating from its initial alignment.
A bow-like structure towards the head portion of the filamentary cloud, could be influenced by local external feedback, internal feedback or gravitational collapse.


\subsection{Dust grain alignment}
\begin{figure}
    \includegraphics[width=1
    \linewidth]{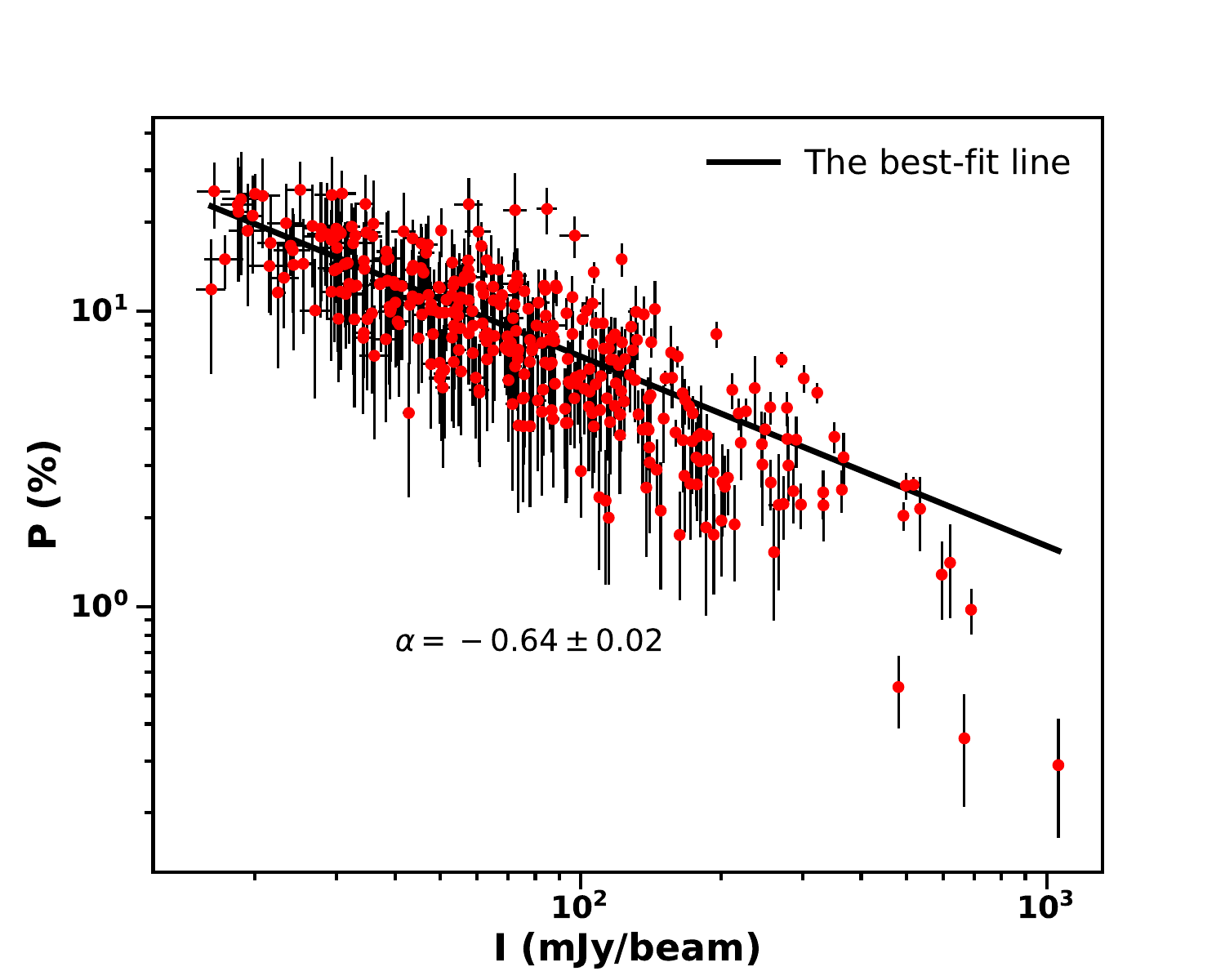}
    \caption{Polarization fraction as a function of total intensity at 850 $\mu$m. Single power law is fitted to the data points.}
    \label{fig:dustalignment}
\end{figure}

The fundamental assumption that dust grains align with B-fields allows us to trace those fields using polarization data. In our analysis, we have considered the debiased polarization fraction (P) as a function of total intensity (I) at 850 $\mu$m modeled using the single power law. If the dust grains are not aligned with the B-field in a dense cloud, the observed polarization intensity (PI) would be independent of the cloud's column density, leading to a relationship between PI and I where P = PI/I is proportional to I$^{-1}$. On the other hand, if B-field aligned dust grains are present in a dense cloud, we expect to see a decrease in polarization fraction (P) as the cloud's column density increases, resulting in a relationship between P and intensity (I) where P is proportional to I$^{-\alpha}$, where $\alpha$ lies between 0 and 1. An index
$\alpha$ $\sim$ 1 indicates a lack of alignment between dust grains and
B-ﬁelds. As illustrated in Figure \ref{fig:dustalignment}, we fitted a single power law to all of the debiased data points. We obtained the best-fit power-law exponent as -0.64$\pm$ 0.02, which shows that the dust grains are aligned with respect to the B-field, and hence our data infers B-field orientation in all the intensities of the region.

\section{Analysis} \label{analysis}
\subsection{Filament width} \label{filament width}
\begin{figure}
    \centering
    \includegraphics[width=1\linewidth]{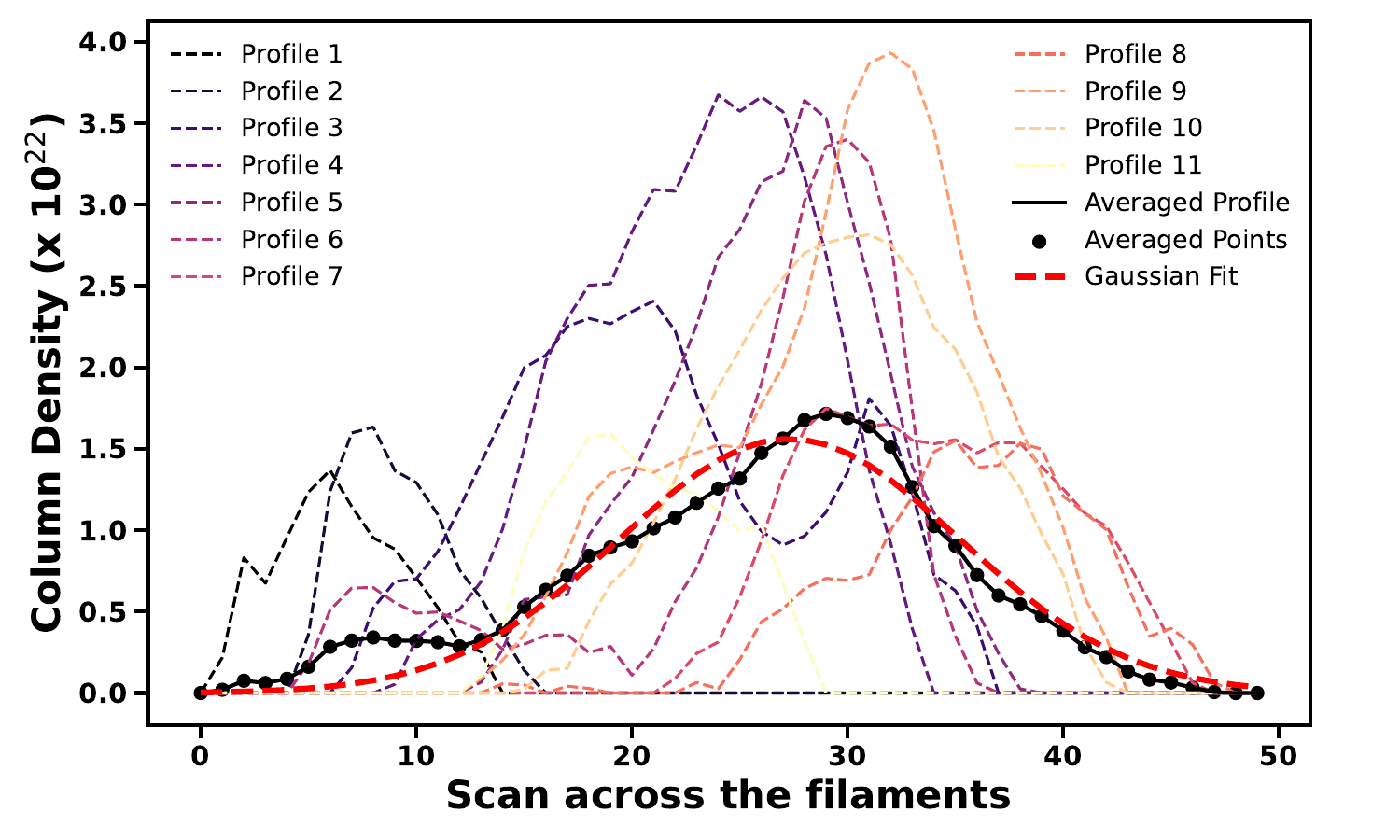}
    \caption{The H$_2$ column density profiles for 11 cuts as shown in Figure \ref{fig:Bfieldmorphology} are shown as dashed lines. The average profile, marked with black circles, is represented by the solid black line, while the fitted Gaussian is overlaid as a red dashed line.}
    \label{fig:filamentwidth}
\end{figure}

To determine the width of the filament, we analyzed the distribution of H$_2$ column density. Profiles of the column density were constructed along eleven cuts oriented perpendicularly to the filament's long axis as shown in Figure\ref{fig:Bfieldmorphology}. These profiles were extracted by sampling the column density data across the filament at evenly spaced intervals. Once these profiles were obtained, they were averaged to create a single representative profile. To quantify the filament's width, we fitted a Gaussian model to the averaged column density profile as shown in Figure \ref{fig:filamentwidth}. The key parameter derived from the fit is the Full Width at Half Maximum (FWHM), which corresponds to the width of the filament at the point where the column density falls to half of its maximum value. From the Gaussian fit, we determined the FWHM of the filament to be 0.070 $\pm$ 0.001 pc. This value represents the physical width of the filament in parsecs and includes the associated fit uncertainty.
\subsection{Velocity dispersion} \label{Velocity dispersion}
In the Cep B region, $^{13}$CO is used to trace moderately dense gas, outlining the cloud structure and revealing large-scale kinematics. While C$^{18}$O is better suited for probing dense cores, $^{13}$CO more effectively captures the extended molecular environment, making it ideal for studying cloud and feedback interactions in star-forming regions like Cep B. To derive the non-thermal velocity dispersion map ($\Delta V_\text{NT}$), we used $^{13}$CO (J = 3–2) data from JCMT/HARP. A single Gaussian profile was fitted to the spectrum at each pixel, with the resulting $\Delta V_\text{total}$ representing the observed full width at half maximum (FWHM) of the line. The non-thermal component was then estimated by subtracting the thermal contribution from the measured FWHM at each pixel, 

\begin{equation}
    \hspace{6em} \Delta V_\text{NT}^2 = \Delta V_{\text{total}}^2 - \frac{k T_\mathrm{k}}{m_{\mathrm{^{13}CO}}} \, 8 \ln 2
\end{equation}

\begin{figure*}
\resizebox{9cm}{6cm}{\includegraphics{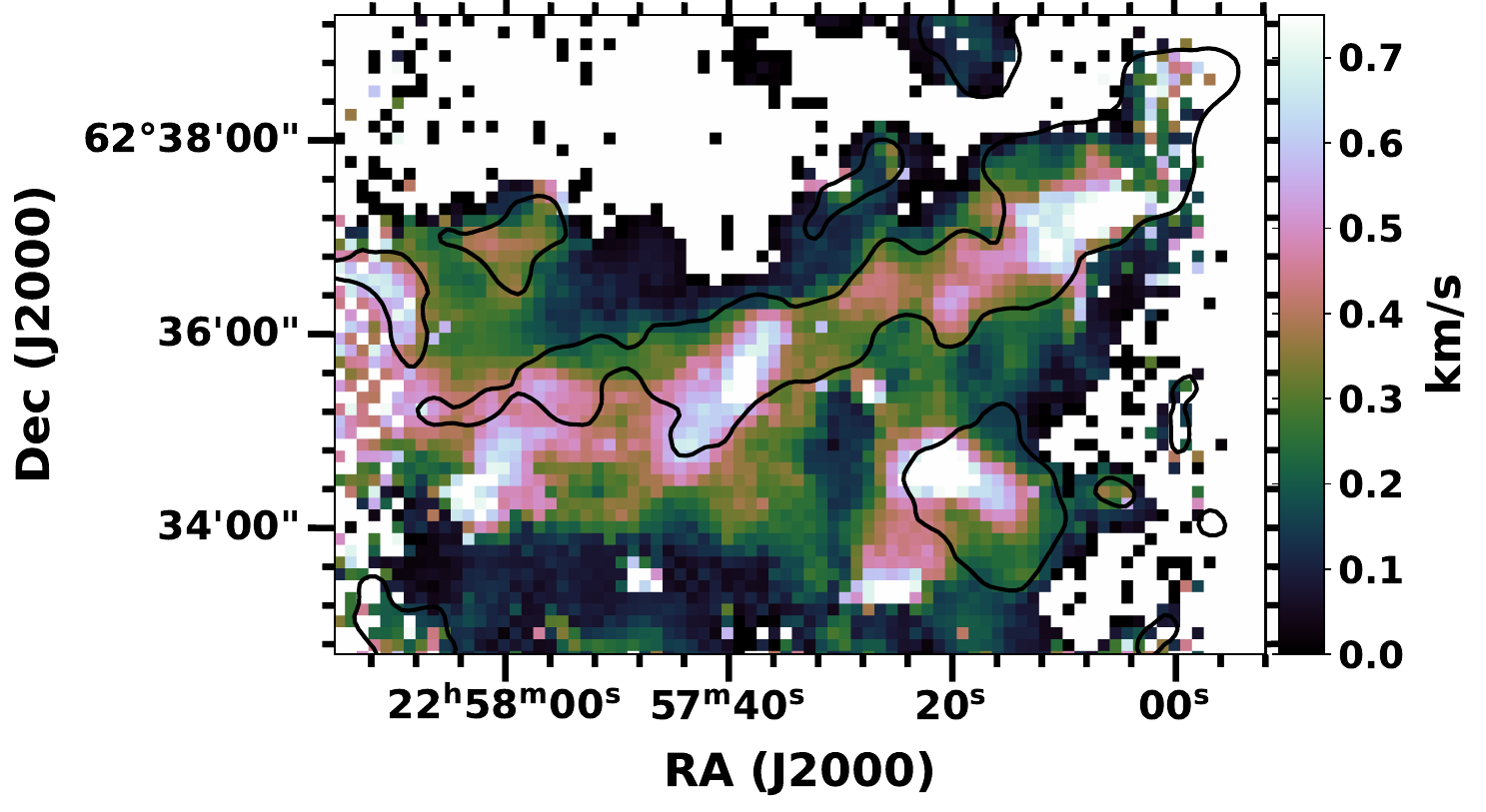}}
\resizebox{7cm}{6cm}{\includegraphics{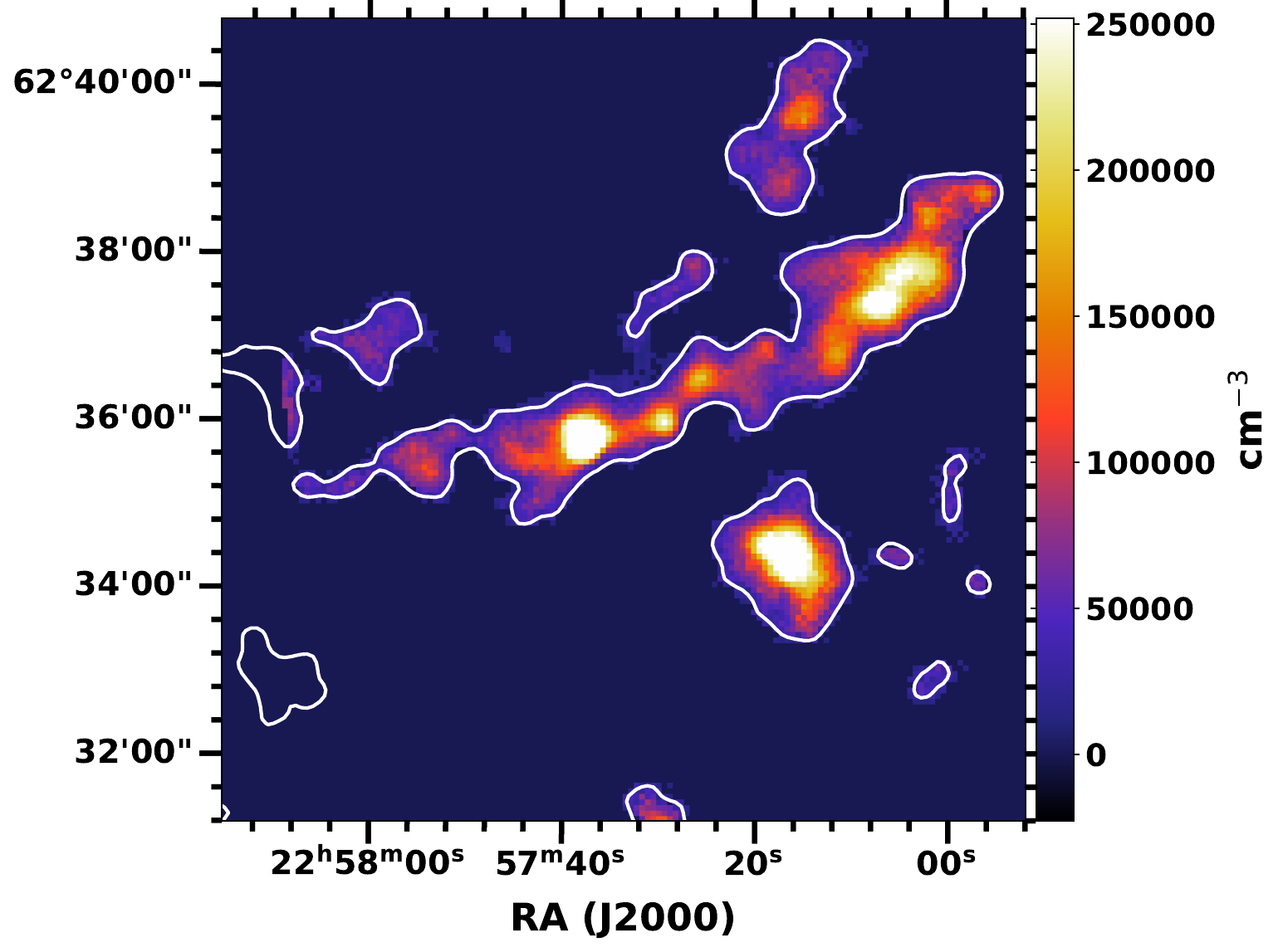}}
\caption{(Left): Velocity dispersion map using the $\mathrm{^{13}CO}$ data, (Right): Volume density map. The contour is same as the contour shown in Figure \ref{fig:Bfieldmorphology_unscaled}}
\label{fig:velocity_dispersion}
\end{figure*}

In this analysis, $\Delta \mathrm{V_{total}}$ denotes the observed FWHM of the $^\mathrm{13}$CO spectral line. The thermal velocity dispersion is given by $\sqrt{\frac{kT\mathrm{_k}}{\mathrm{m_{^{13}CO}}}}$, where T$\mathrm{_k}$ is the kinetic temperature and $\mathrm{m_{^{13}CO}}$ is the mass of the $^\mathrm{13}$CO molecule. We assume that the kinetic temperature of $^\mathrm{13}$CO is equivalent to the dust temperature at each pixel. The dust temperature map was derived through spectral energy distribution (SED) fitting to the Herschel dust continuum images, with further details provided in Appendix \ref{appendix b}. The velocity dispersion map is presented in the left panel of Figure \ref{fig:velocity_dispersion}.
\subsection{Volume density} \label{Volume density}
To estimate the B-field strength, we need an estimate of gas number density. For this, we need to take the ratio between the column density map and the width of the filamentary cloud. Here we obtain the Herschel column density map using the Spectral Energy Distribution (SED) fitting with $\it{Herschel}$ PACS 160 $\mu$m, SPIRE 250 $\mu$m, 350 $\mu$m, 500 $\mu$m and JCMT/SCUBA-2 850 $\mu$m data (see Appendix \ref{appendix b}). To get the volume density map, we divided the column density map by the width of the cloud. It is quite difficult to get the three-dimensional information from dust polarization and spectral line observations. To get around this issue, researchers assume that integrated polarized emission arises from a particular depth along the line of sight at which they assume an effective volume density. Then they measured the B-field strength in the plane perpendicular to the sight line at depth. In this study, we assumed that the volume density in the Cep B region is proportional to the column density in order to achieve an effective volume density. Assuming a constant depth over the area of interest is the easiest way to express proportionality. We calculated the width of the filamentary cloud as explained in \ref{filament width}. Then the volume density is calculated using the following relation:

\begin{equation}
    \hspace{6em} n(\mathrm{H}_2) = \frac{N(\mathrm{H}_2)}{W}.
    \label{eq:volden}
\end{equation}

The obtained volume density map is shown in the right panel of Figure \ref{fig:velocity_dispersion}.
\subsection{Structure Function (SF) analysis} \label{st_func}
In the structure function (SF) analysis \citep{hildebrand2009dispersion}, the B-field is modeled as the sum of a large-scale ordered field (B$_{\mathrm{0}}$) and a turbulent component $\delta$B. The SF quantifies how position angle dispersion varies with vector separation `$\text l$'. At scales larger than the turbulence correlation length $\delta$, $\delta$B should reach its maximum. At smaller scales ($l$ < $d$), higher-order terms in the Taylor expansion of 
B$_{\mathrm{0}}$ can be ignored. When $\delta$ < $l$ $\ll$ d, the SF follows a characteristic form,

\begin{equation*}
    \hspace{6em} \langle \Delta \phi ^ 2 (l)  \rangle _\mathrm{tot} - \sigma^2_\mathrm{M} (l) \simeq b^2 + m^2 l^2 .
\end{equation*}

In this equation, $\langle \Delta \phi ^ 2 (l)  \rangle _{tot}$, the square of the total measured dispersion function, comprises three components: $b^2$, a constant contribution from turbulence; $m^2l^2$, the contribution from the large-scale structured B-field; and $\sigma^2_M$ ($l$), the contribution from measurement uncertainty. The ratio of the turbulent component to the large-scale component of  B-field is expressed as
\begin{equation}
    \hspace{6em} \frac{\langle \Delta B ^ 2 \rangle ^{1/2}} {B_0} = \frac{b}{\sqrt{(2-b^2)}} ,
\end{equation}
where, B$_{\mathrm{0}}$ is the strength obtained by the modified Davis-Chandrasekhar-Fermi (DCF) relation,
\begin{equation}
    \hspace{6em} {B_0} \approx \sqrt{(2-b^2) 4\pi \mu m_H n_{H_2}} \frac {\sigma_v} {b}.
    \label{eq:modified_dcf}  
\end{equation}
Here, n$_\mathrm{H_2}$ denotes the average volume density, and $\sigma_v$ represents the velocity dispersion of the $^{13}$CO molecular line, as described in Sections \ref{Volume density} and \ref{Velocity dispersion}, respectively. We adopt a average value of 1.06 $\times$ 10$^5$ cm$^{-3}$ for n$_\mathrm{H_2}$ and 0.65 km s$^{-1}$ for $\sigma_v$ within the filament. The estimated B field strength in the plane of the sky is then calculated by applying a correction factor Q
\begin{equation}
    \hspace{6em} B{\mathrm{_{POS}}} = Q B_0.
\end{equation}
The factor Q is taken to be 0.5, based on studies using synthetic polarization maps generated from numerical simulations of clouds \citep{ostriker2001density}. These studies indicate that the B-field strength may have an uncertainty of up to a factor of 2 when the B-field dispersion is $<= 25 ^{\circ}$.

\begin{figure*}
    \centering
    \includegraphics[width=0.5\textwidth]{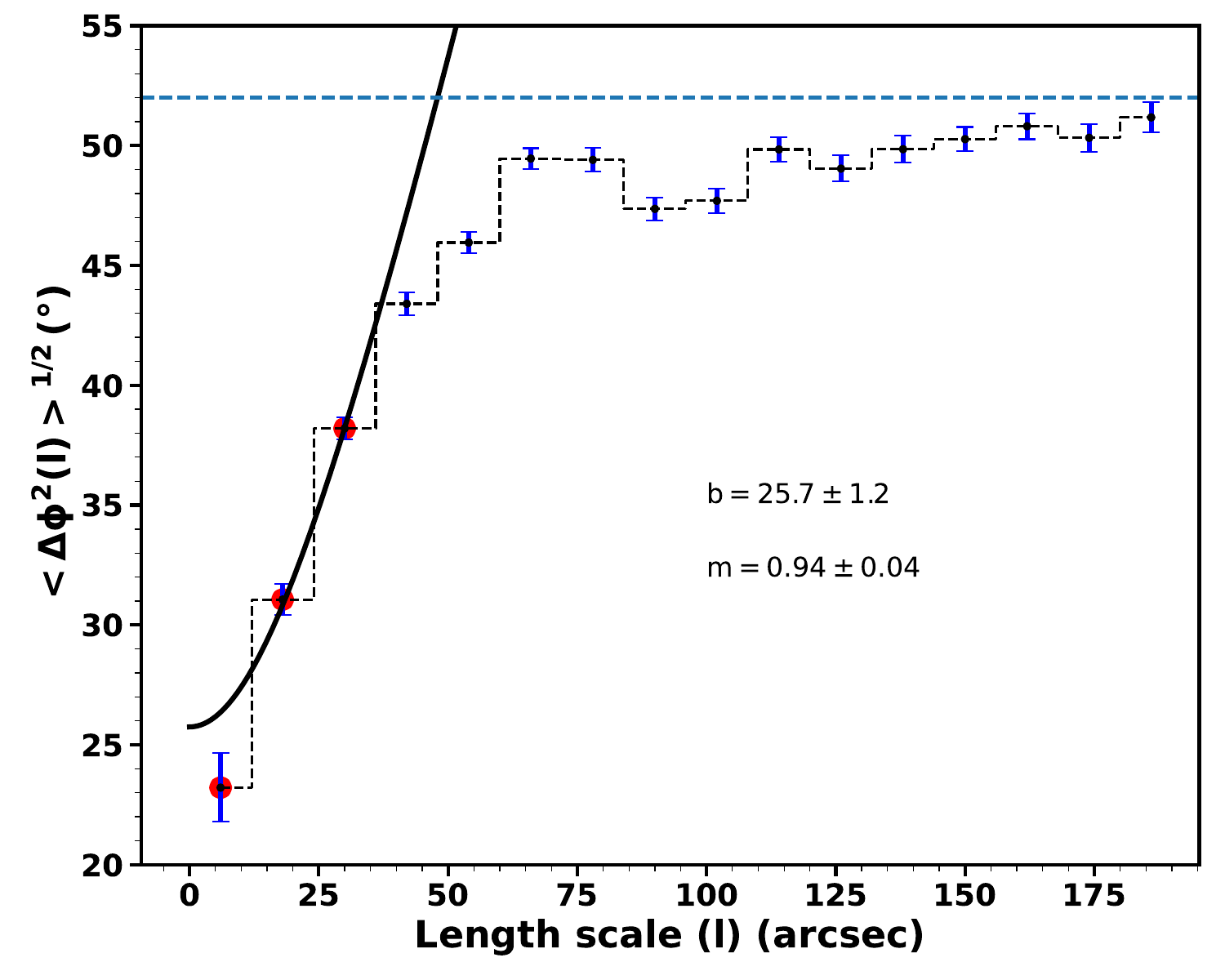}
    \caption{Angular dispersion function as a function of length scale for the filamentary structure. The plotted angular dispersions (red filled circles) have been corrected for measurement uncertainties. The best-fit line is shown in black, with the data points used in the fits indicated by circled symbols. The blue dashed line is drawn at 52$^\circ$.}    
    \label{fig:SF}
\end{figure*}
The Figure \ref{fig:SF} represents the angular dispersion corrected for uncertainties $(\langle \Delta \phi ^ 2 (l)  \rangle _{tot} - \sigma^2_M (l))$ as a function of length scale, measured from the POL-2 dust polarization data. The peak value in the current structure function is lower than the 52$^\circ$
 expected for a completely random field \citep{poidevin2010magnetic}. The red fitted circles represent the data points for which the nonlinear least-squares fitting is performed. The fitting parameter `b' is obtained to be 25.6 $^\circ$ $\pm$ 0.9 $^\circ$. The resulting $\langle \Delta B ^ 2 \rangle ^{1/2}$ /B$_{\mathrm{0}}$ value is 0.33 $\pm$ 0.01 for Cep B by considering the B-field segments only along the filamentary structure. We derived the B-field strength as 181 $\pm$ 9 $\mu$G by using the modified DCF relation as in Equation \ref{eq:modified_dcf}.
\subsection{Autocorrelation Function (ACF) Analysis}
The ACF analysis \citep{houde2009dispersion}, extends the SF analysis by accounting for the effects of signal integration both along the line of sight and within the beam. As described by \citet{houde2009dispersion}, the ACF can be expressed as
\begin{align}
    \hspace{2em} 1 - \left\langle \cos\left[\Delta \Phi(l)\right] \right\rangle \\
    &\approx \frac{1}{N} \frac{\left\langle \delta B^2 \right\rangle}{\left\langle B_0^2 \right\rangle}
    \times \left[1 - \exp\left(-\frac{l^2}{2(\delta^2 + 2w^2)}\right)\right] + a_2' l^2,
    \label{eq:acf}
\end{align}
where, $\Delta \phi$ ($l$) represents the difference in position angles between two vectors separated by a distance $l$. $W$ is the beam radius, which for the JCMT is 6 $\arcsec$ (calculated as the FWHM beam size of 14$\arcsec$ divided by 8 $\ln$ 2). $a_2^{\prime}$ is the slope of the second-order term in the Taylor expansion, while $\delta$ is the turbulent correlation length. $N$ represents the number of turbulent cells within the telescope beam, and it is defined as
\begin{equation*}
    \hspace{6em} \mathrm{N} = \frac{(\mathrm{\delta}^2 + 2 \, \mathrm{w}^2) \, \mathrm{\Delta}'} {\sqrt{2\pi} \, \mathrm{\delta}^3} ,
\end{equation*}
here $\Delta$ is the effective thickness of the cloud. The ordered B-field strength is given as
\begin{align}
    \hspace{6em} \mathrm{B}_0 \approx \sqrt{4\pi \, \mu \, m_\mathrm{H} \, n_{\mathrm{H}_2}} \, \sigma_v 
    \left[\frac{\langle \delta \mathrm{B}^2 \rangle}{\mathrm{B}_0^2}\right]^{-1/2} .
    \label{eq:mod_dcf}
\end{align}

\begin{figure*}

\resizebox{8cm}{6cm}{\includegraphics{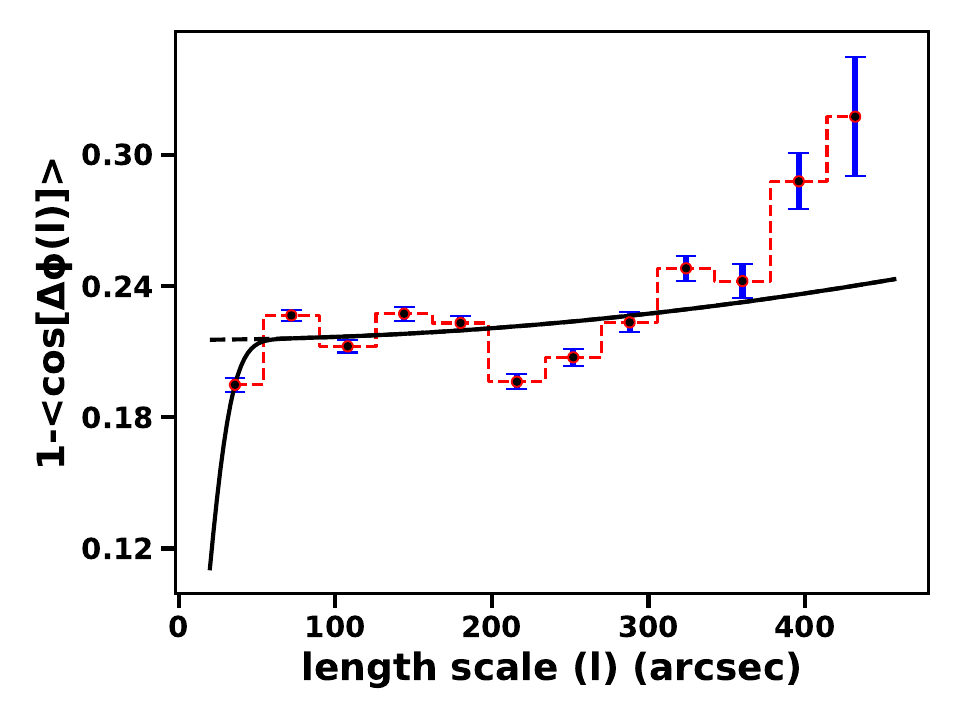}}
\resizebox{8cm}{6cm}{\includegraphics{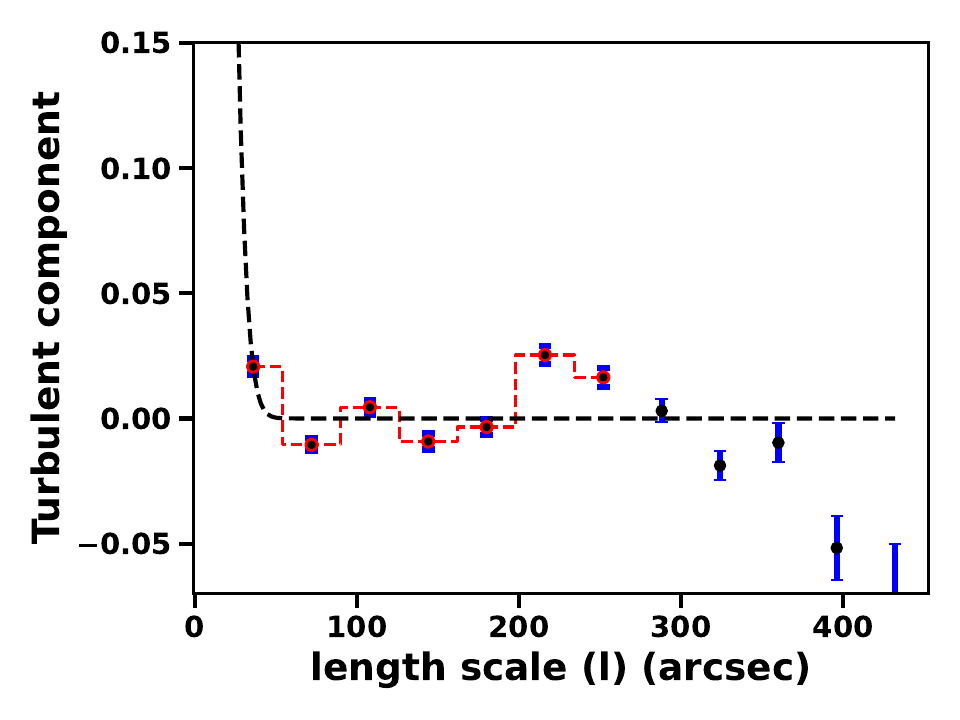}}
\caption{(Left): Autocorrelation function as a function of length scale for the filamentary structure. Angular dispersions are represented by filled circles. The best-fit dispersion function is shown as a solid black line, while the red dashed line represents the contribution from large-scale ordered or uncorrelated components, modeled as $\frac{1}{N} \frac{\langle \delta B^2 \rangle}{\langle B_0^2 \rangle}+ a_2' l^2$. Data points used in the fitting are highlighted with encircled filled symbols. (Right): The best-fit model for the turbulent or correlated component, given by $\frac{1}{N} \frac{\langle \delta B^2 \rangle}{\langle B_0^2 \rangle} \times e^{-l^2 / 2(\delta^2 + 2w^2)}$, is shown as dashed black lines. The plotted data points represent the residual angular dispersions—i.e., the difference between the observed values (encircled filled symbols in the left panel) and the modeled large-scale ordered component (black dashed line in the left panel).}
\label{fig:Autocor}
\end{figure*}
Equation \ref{eq:acf} is fitted to the ACF data as shown in Figure \ref{fig:Autocor} for the filament. The bin width for constructing the ACF (1-$\langle \cos[\Delta \Phi (l)] \rangle$) was chosen to be 36$\arcsec$. Various bin widths were investigated; we found that best fit was achieved with 36$\arcsec$. Using the fitted parameter ⟨$\delta$B$^2$⟩/⟨B$_0^2$⟩ as 0.47$\pm$0.02, along with derived parameters such as number densities and velocity dispersions, we have estimated the B-field strength using the modified DCF relation (Equation \ref{eq:mod_dcf}). The value of N$_\mathrm{H_2}$ and $\sigma_v$ is same as mentioned in section \ref{st_func}. The estimated B-field strength is 153 $\pm$ 6 $\mu$G in the filament.
\vskip 4pt

\subsection{Polarization angle dispersion} \label{Polarization angle dispersion}
Estimating the mean B-field orientation is essential for accurately quantifying the turbulent dispersion of B-field lines. However, determining the local mean field orientation from polarization observations is challenging due to the distortion of B-field lines in a molecular cloud caused by turbulence and gravity. To determine the angle dispersion at each pixel, we utilized a method proposed by \cite{hwang2021jcmt}. This involved selecting a small box with dimensions of $36\arcsec \times 36\arcsec$ and centered at a pixel and then computing the root mean square of angle differences. This value represented the polarization angle dispersion at the central pixel of the box. We repeated this process by moving the box and estimating angle dispersion at each pixel. 

To derive the B-field strength map using the DCF method, we have used the maps of velocity dispersion, volume density, and polarization angle dispersion, as mentioned in sections \ref{Velocity dispersion}, \ref{Volume density} and \ref{Polarization angle dispersion} respectively.
\subsection{DCF analysis to obtain the B-field strength map}
\begin{figure}
    \centering
    \includegraphics[width=\linewidth]{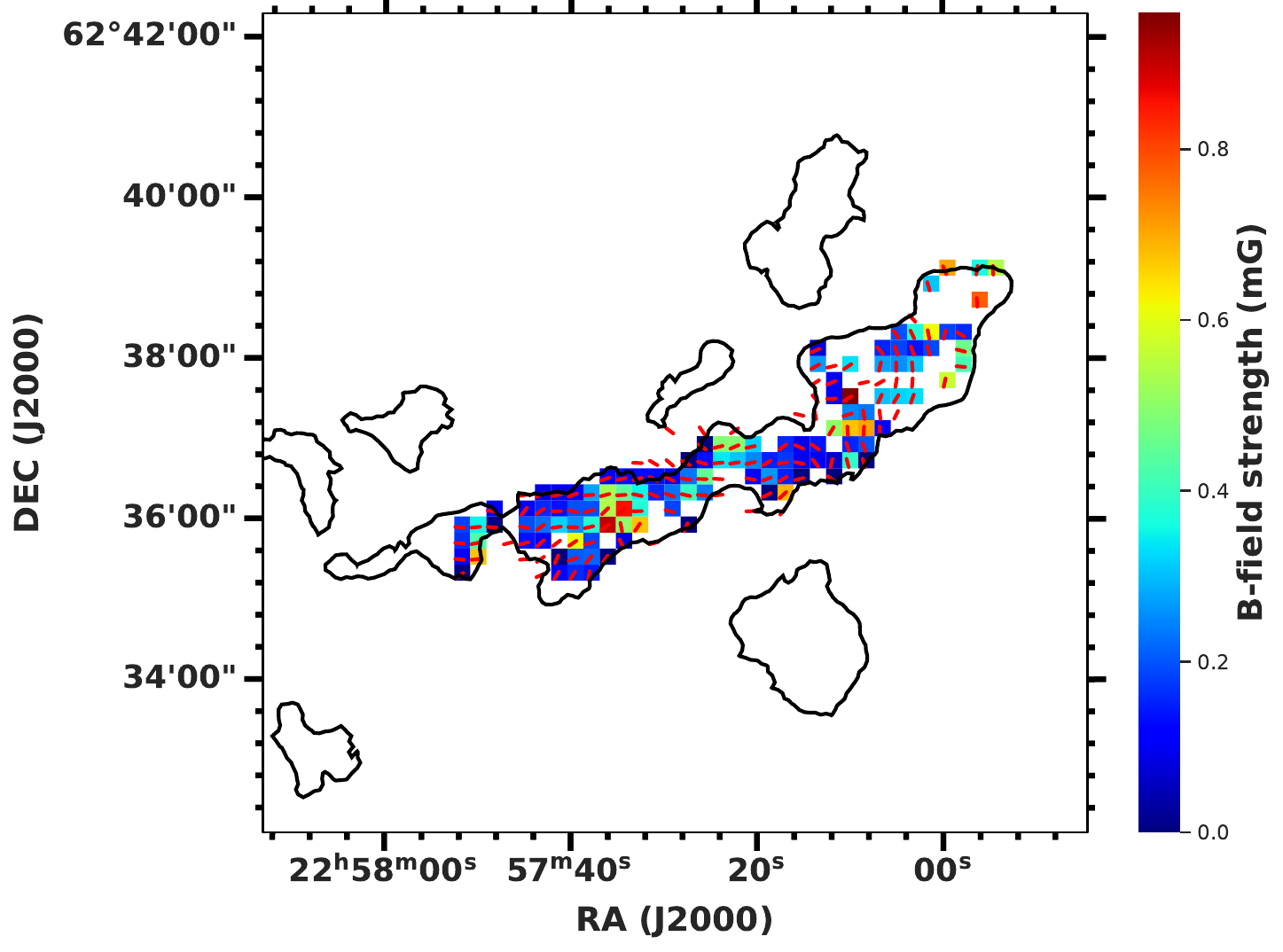}
    \caption{B-field strength map estimated using the Davis–Chandrasekhar–Fermi (DCF) method. The black contours is same as shown in Figure \ref{fig:Bfieldmorphology_unscaled}}
    \label{fig:DCF}
\end{figure}

The DCF method \citep{davis1951strength,chandrasekhar1953problems} has been used to estimate the B-field strengths in Cep B star-forming region. This method assumes that the perturbations in the B-field are Alfv{\'e}nic which means the deviation in angle from the mean field direction is due to the distortion by small scale non-thermal motions. Using this method, we generate a map of the strength of the B-field (B-field) by utilizing these three observed quantities: velocity dispersion, gas number density and polarization angle dispersion in sections \ref{Velocity dispersion}, \ref{Volume density} and \ref{Polarization angle dispersion} respectively. To obtain the B-field strength map, we acquired these three separate maps. We then interpolated the volume density and velocity dispersion maps to align their pixel values with the coordinates of the angle dispersion map. Then, we inserted the values of these three quantities and obtained the plane-of-sky B-field strength (B$_\mathrm{pos}$) using the DCF relation \citep{crutcher2004drives}

\begin{align}
    B_{\mathrm{DCF}}\, [\mu\mathrm{G}] 
    &= Q \sqrt{4\pi \rho} \frac{\sigma_v}{\sigma_\theta} \\
    &\approx 9.3\, \sqrt{n(\mathrm{H}_2)\, [\mathrm{cm}^{-3}]} 
    \frac{\Delta V_\text{NT}\, [\mathrm{km\,s}^{-1}]}{\sigma_\theta\, [\mathrm{deg}]},
    \label{eq:dcf}
\end{align}

where Q=0.5 is the correction factor suggested by \cite{ostriker2001density} in the case where angle dispersion is less than 25$^\circ$, $\rho$ is the gas density, $\sigma_v$ is the non-thermal velocity dispersion, and $\sigma_{\theta}$ represents the angular dispersion of the plane-of-sky component. In the above equation \ref{eq:dcf}, n($\mathrm{H_2}$)= $\rho$/$\mu \mathrm{m_{H}}$ and FWHM non-thermal velocity dispersion $\Delta V_{\mathrm{NT}}$ = $\sigma_{\mathrm{NT}}$ $\sqrt{8ln2}$ and assumes a molecular weight $\mu$=2.8 \citep{kirk2013first} which we adopted throughout this work. The B-field strength map obtained is shown in Figure \ref{fig:DCF}.

\subsection{Thermal and Radiation pressure in Cep B}\label{rad_thermal_pressure}
\begin{figure}
    \centering
    \includegraphics[width=\linewidth]{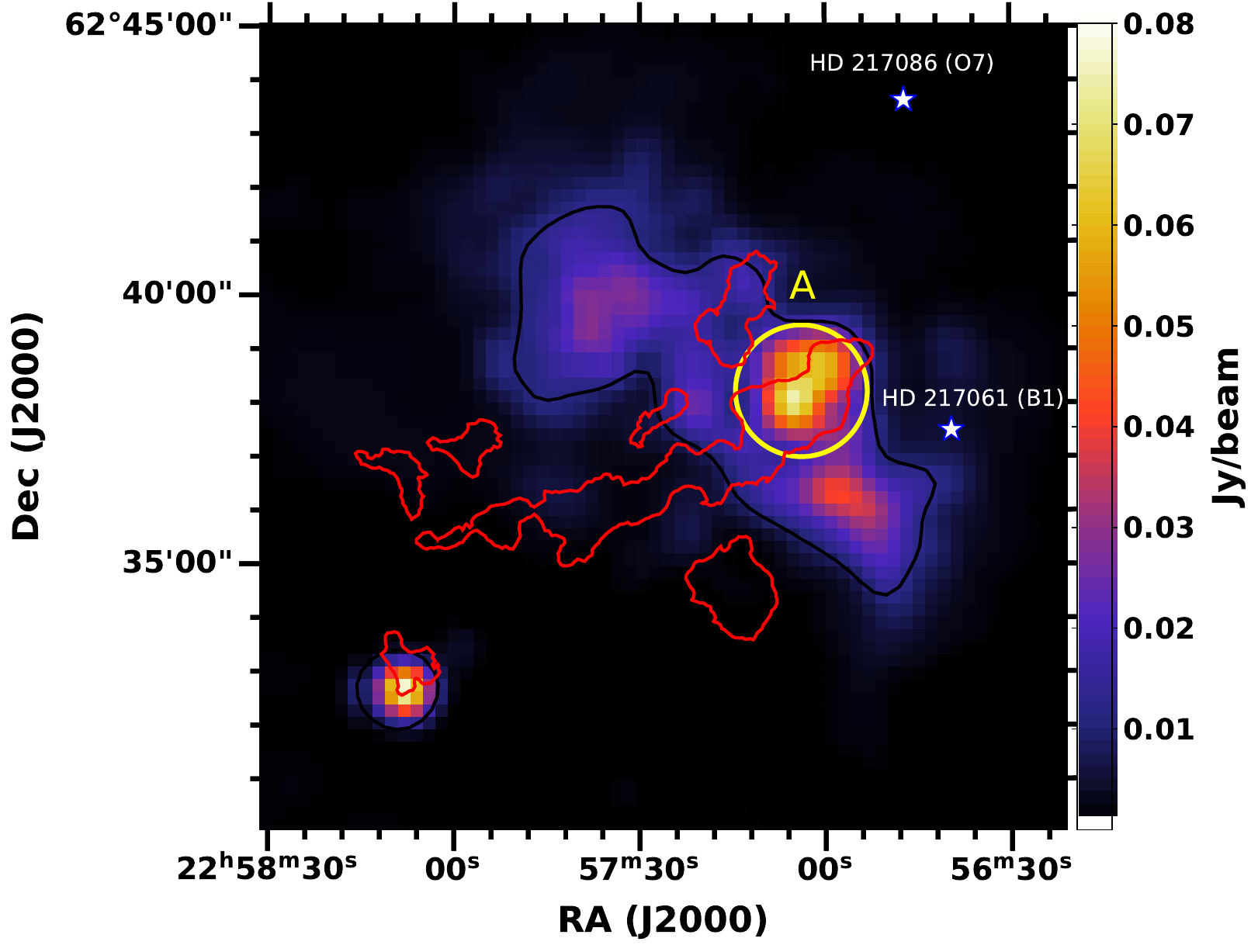}
    \caption{Radio continuum map of the Cep B at 1.4 GHz. The red contour represents the filamentary structure as in Figure \ref{fig:Bfieldmorphology} and the black contour is plotted at 0.01 Jy/beam. The yellow circular region we have considered to calculate the thermal and radiation pressure. The two OB stars are shown by blue stars.}
    \label{fig:VLA}
\end{figure}

The radio continuum view of the Cep B H{\sc ii} region at a wavelength of 21 cm (1.4 GHz), as observed with the VLA, is presented in Figure \ref{fig:VLA}. The peak flux density (S$_\nu$) is 0.240 ± 0.003 Jy, with the peak emission region highlighted in yellow and labeled as `A'. We focused on this region due to the spatial correspondence between the radio data and the dust continuum emission. To estimate the electron density (n$_e$) of the ionized gas, we assume an electron temperature (T$_e$) as 10$^4$K \citep{stahler2005formation}. The electron density was determined using the empirical relation provided by \cite{martin2005high}:

\begin{align} 
\hspace{4em}
    n_e &= \frac{4.092 \times 10^5\, \mathrm{cm}^{-3}}{\sqrt{b(\nu, T_e)}} 
    \left( \frac{S_\nu}{\mathrm{Jy}} \right)^{0.5}
    \left( \frac{T_e}{10^4\, \mathrm{K}} \right)^{0.25} \nonumber \\
    &\quad \times \left( \frac{D}{\mathrm{kpc}} \right)^{-0.5}
    \left( \frac{\theta_D}{\arcsec}\right)^{-1.5},
    \label{eq:ne_density}
\end{align}

where, 
\begin{align}
\hspace{4em}
    b(\nu, T_e) = 1 + 0.3195 \log\left( \frac{T_e}{10^4\, \mathrm{K}} \right)
    - 0.2130 \log\left( \frac{\nu}{\mathrm{GHz}} \right).
    \label{eq:b_correction}
\end{align}

Here $\theta_\text D$ is the angular diameter of the source, D (725 pc) is the distance from the Sun. With the derived electron density, we further calculated the corresponding thermal pressure (P$_\mathrm{Te}$) exerted by the ionized gas using the relation:
\begin{align}
    \hspace{6em}
    P_{\mathrm{Te}} = 2 n_e k_{\mathrm{B}} T_e ,
\end{align}

where k$_B$ is the Boltzmann constant. This yields a thermal pressure of $\approx$ 21.4 $\times$ 10$^{-10}$ $\mathrm{dyn \, cm^{-2}}$.

We estimated the mean radiation pressure (P$_\mathrm{rad}$) driven by ionizing stars of spectral type O7 and B1. The photon emission rates (in photons s$^{-1}$) from O7 and B1 star given by 10$^{49.06}$ and 1.9 $\times$ 10$^{45}$ photons s$^{-1}$ \citep{sternberg2003ionizing,ojha2004study}. The ionizing flux emitted by one O7 star is 11.34 $\times$ 10$^{10}$ photons cm$^{-2}$ s$^{-1}$ and B1 star is 3.18 $\times$ 10$^{8}$ photons cm$^{-2}$ s$^{-1}$. Each UV photon carries an energy of h$\nu$ = 20 ev, so estimated the radiation pressure using the relation P$_\mathrm{rad}$ = h$\nu$ q$_0$/c as 1.2 $\times$ 10$^{-10}$ dyn cm$^{-2}$ from O7 star and 0.34 $\times$ 10$^{-12}$ dyn cm$^{-2}$ from B1 star respectively.

\subsection{Histogram of Relative Orientation (HRO)} \label{HRO}
The Histogram of Relative Orientations (HRO) is widely used in both numerical simulations and observational studies to assess the alignment between density or column density structures and the local B-field \citep{soler2013,ade2016planck,kwon2022b}. The relationship between B-fields and filamentary structures can be quantitatively analyzed using HRO method. This technique compares the orientation of polarization vectors ($\hat{E}$) with the gradients of optical depth ($\nabla \tau$). Specifically, the HRO involves constructing a histogram of the relative angles ($\phi$) between $\hat{E}$ and $\nabla \tau$ at various density regimes. In this study, we adopt a simplified approach by using intensity gradients instead of optical depth gradients.

The histogram shape parameter ($\xi$) is described
by the parameter
\begin{align}
\centering
    \hspace{6em} \xi =& \frac{A_c - A_e}{A_c + A_e},
\end{align}

where, the angle $\phi$ in our analysis ranges from 0$^\circ$ to 90$^\circ$. A$_c$ denotes the area in the histogram concentrated near 0$^\circ$ (0 $^\circ$ < $\phi$ < 22.5 $^\circ$), while A$_e$ corresponds to the area concentrated near 90$^\circ$ (70$^\circ$ < $\phi$ < 90$^\circ$). When the relative orientation angles are small ~--~ meaning polarization vectors are mostly aligned with intensity (or optical depth) gradients ~--~ the histogram shape parameter $\xi$ takes on positive values because $A_c > A_e$. Conversely, if the polarization is primarily perpendicular to the intensity gradients, the $\xi$ becomes negative. Given that polarization is orthogonal to the B-field and filaments are oriented perpendicular to intensity gradients, a positive $\xi$ indicates that B-fields are aligned with filaments, while a negative $\xi$ suggests that B-fields are oriented perpendicular to them. The uncertainties in $\xi$ are primarily influenced by the number of data points within each histogram bin and are computed following the method outlined in \citep{ade2016planck}.

The corresponding uncertainty $\sigma_\xi$ is expressed as
\begin{align}
    \hspace{6em} \sigma_\xi =& \frac{4(A_e^2 \sigma^2_{A_c} - A_c^2 \sigma^2_{A_e})}{(A_c + A_e)^4}.
\end{align}
The variances $\sigma^2_{A_c}$ and $\sigma^2_{A_e}$ of A$_c$ and A$_e$ are
\begin{align}
    \hspace{6em} \sigma^2_{A_{c,e}}= h_\mathrm{k}(1 - h_\mathrm{k}/h_\mathrm{tot}).
\end{align}
Here, h$_\mathrm{k}$ represents the number of data points in the central or extreme bins, while h$_\mathrm{tot}$ denotes the total number of data points.

\begin{figure}
\includegraphics[width=1\linewidth]{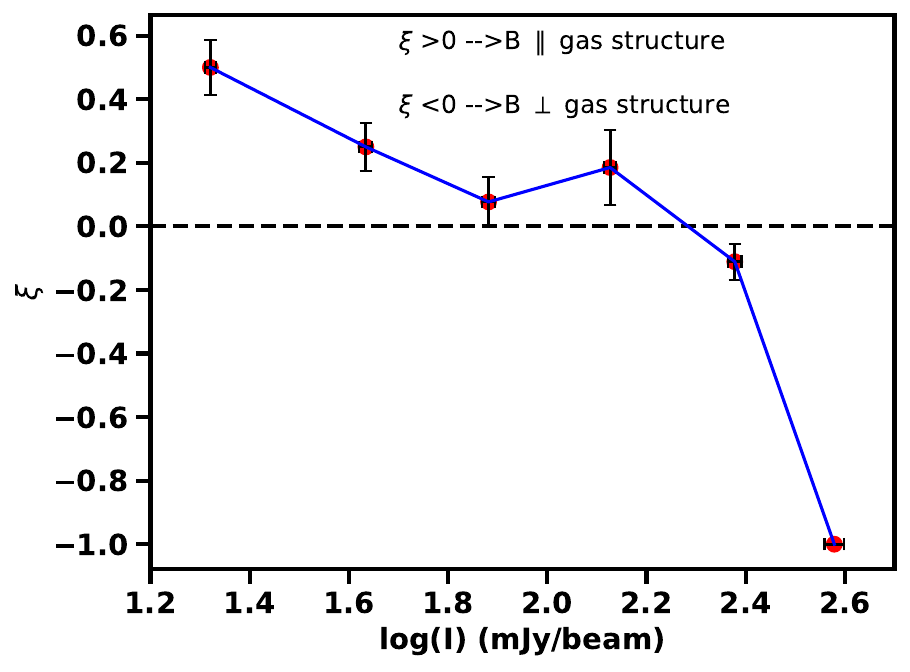}
\caption{The histogram shape parameter ($\xi$) by considering the intensity gradients and B-field PAs (Left), column density gradients and B-field PAs (Right)}
\label{fig:hro}
\end{figure}

Figure~\ref{fig:hro} presents the histogram shape parameter $\xi$ derived using both the dust continuum intensity gradients 
to assess the relative orientation between the B-field and filamentary structures. In this figure, we observe a clear transition in the sign of $\xi$ ~--~ from positive to negative ~--~ as the intensity increases. This transition signifies a change in the relative alignment between the B-field and the gas structure. At lower intensities, the positive values of $\xi$ indicate that the B-field is predominantly aligned parallel to the filamentary structures. However, as the density increases beyond a certain threshold, the shift to negative $\xi$ values suggests that the B-field becomes preferentially oriented perpendicular to the dense structures in the Cep B filament. 

\subsection{Local Gravitational Field}
The local gravitational field plays a crucial role in shaping the formation and evolution of stars within star-forming regions. The spatial distribution of mass directly influences the gravitational potential, which in turn governs the motion and dynamics of gas and dust. Gravitational forces can drive the collapse of material, leading to the development of dense cores and eventually star formation. Therefore, characterizing the local gravitational field is essential for understanding star formation mechanisms and the broader evolution of such regions. This can be achieved through observations that trace the spatial distribution of gas and dust.

To understand the role of gravity in our target star-forming region, we computed the projected gravitational vector field based on JCMT 850 $\mu$m continuum emission. At each map pixel, the direction and relative magnitude of the gravitational field were estimated by summing the dust emission contributions from surrounding pixels. This vector summation accounts for both the direction and distance of neighboring pixels, with each contribution weighted by its dust emission intensity. A key assumption in this method is that the dust emission distribution closely traces the total mass distribution, thereby allowing the dust continuum to serve as a proxy for mapping the gravitational field. Although the gas-to-dust mass ratio can be used to scale the absolute strength of the gravitational vectors, it does not affect their direction.

The local projected gravitational force at a given pixel, denoted as F$_\mathrm{G,i}$, can be estimated using the polarization–intensity gradient method introduced by Koch et al. \citep{koch2012b, koch2012a}. This approach involves calculating the cumulative gravitational influence from all neighboring pixels, expressed as a vector sum of the individual gravitational components,

\begin{center}
\begin{align}
    \hspace{6em} \vec{F}_{\mathrm{G},i} &= \mathrm{K} \, I_\mathrm{i} \sum_{j=1}^{n} \frac{I_\mathrm{j}}{r_\mathrm{ij}^2} \, \hat{r}_\mathrm{ij}
\end{align}
\end{center}

\begin{figure*}
\resizebox{8cm}{6cm}{\includegraphics{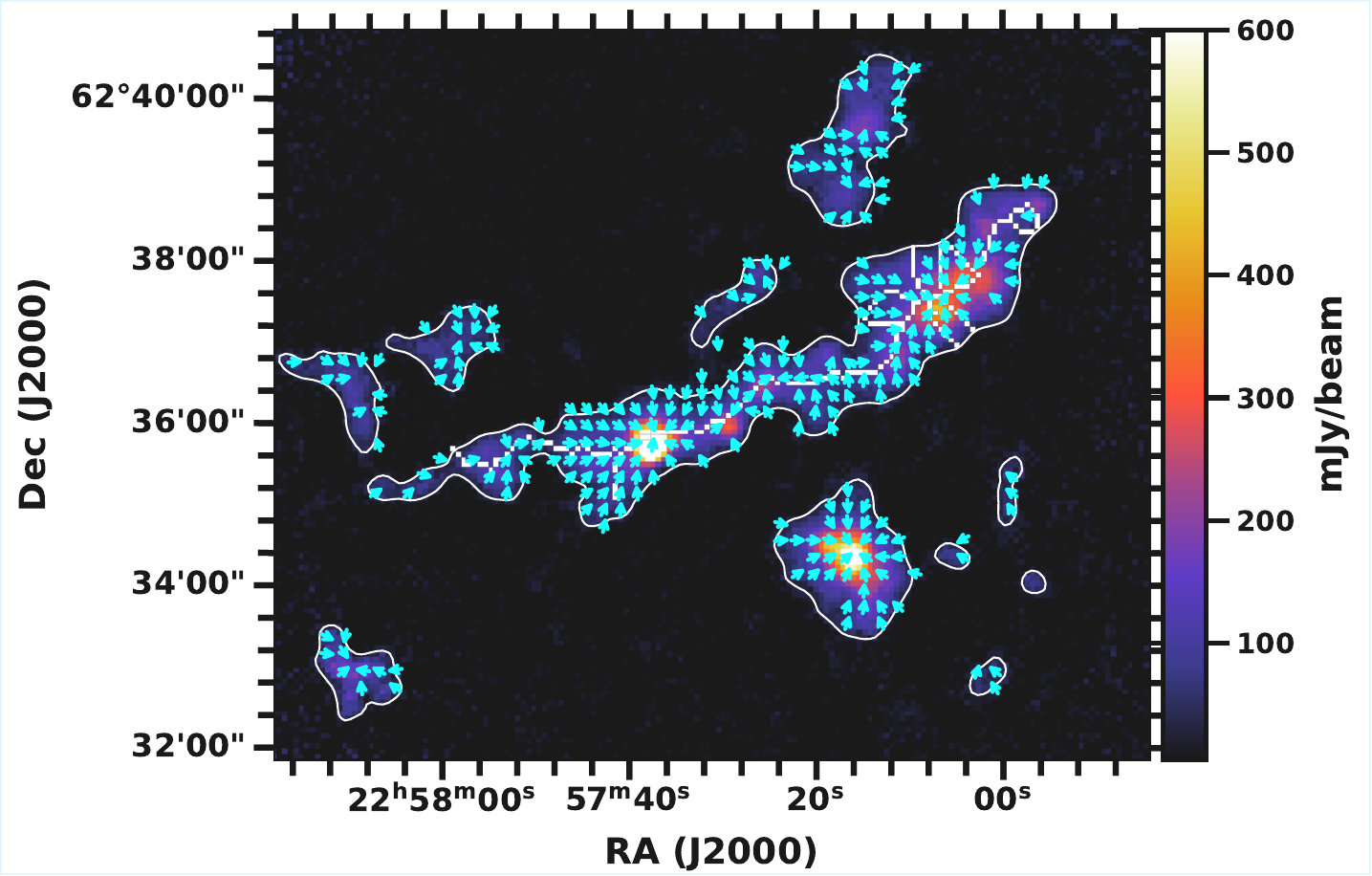}}
\resizebox{8cm}{6cm}{\includegraphics{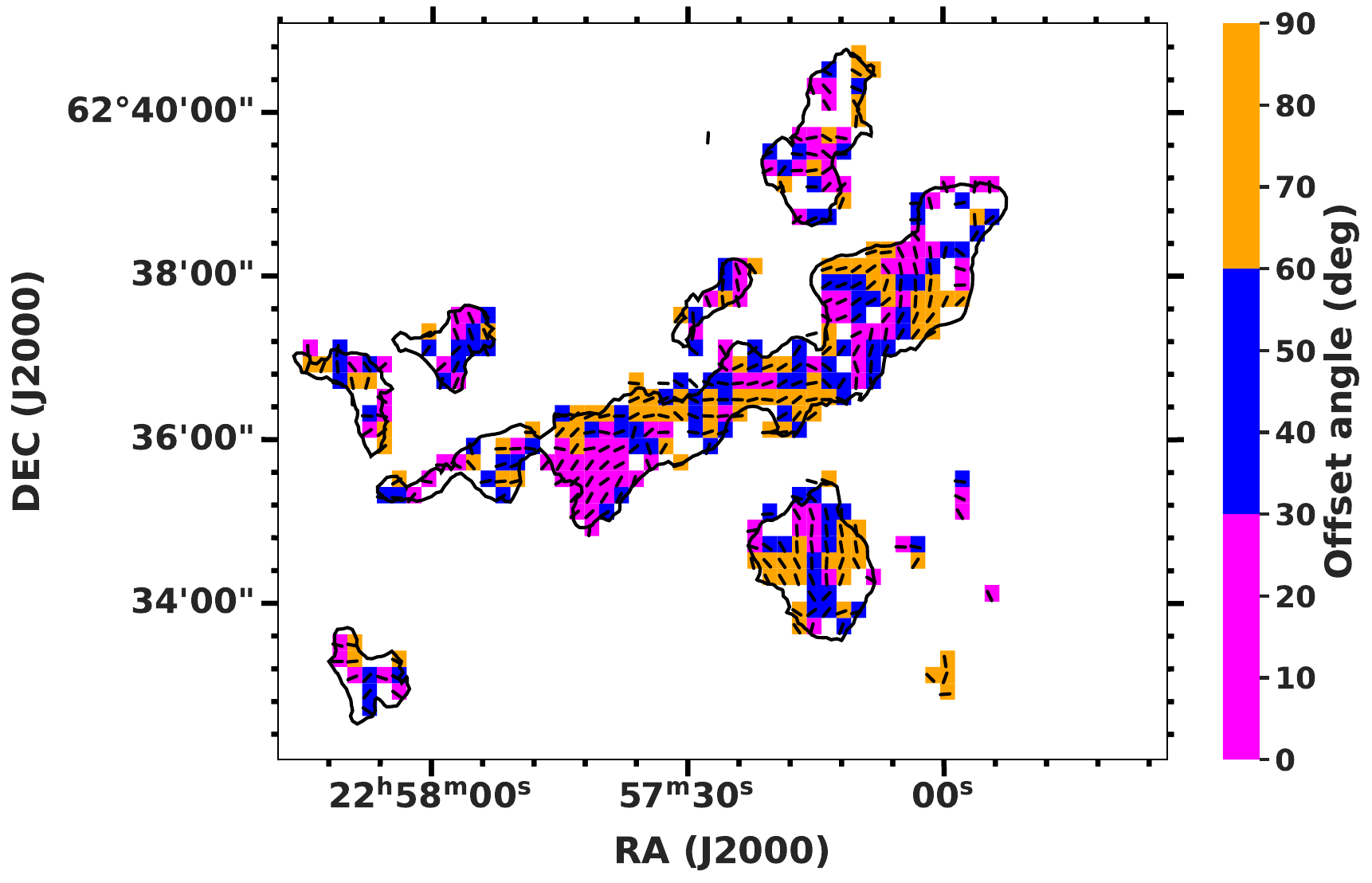}}
\caption{(Left): The projected gravitational field vector map of Cep B. The white skeleton represents the filament idenified from the FILFINDER algorithm. (Right): Represents the offset map between the Gravity vector and the B-field PA}
\label{fig:gravity_vector}
\end{figure*}

The local projected gravitational force at a given pixel, $F_\mathrm{G,i}$, is calculated using an expression that relates dust emission to total column density. This calculation involves the intensities at pixels $i$ and $j$ ($I_\mathrm{i}$ and $I_\mathrm{j}$), the number of contributing pixels $n$, and the projected distance between them on the plane of the sky $r_\mathrm{i,j}$. The corresponding unit vector $\hat{r}_\mathrm{ij}$ helps determine the direction of the local gravitational pull at each pixel. In this analysis, we focus solely on the direction of gravitational forces, under the assumption that the distribution of dust emission effectively traces the overall mass distribution. The constant K is set to 1 for simplicity, and a lower intensity threshold is applied to filter out weak or symmetrically diffuse emission that may not contribute significantly to the local gravitational field. Emission from regions beyond the spatial limits of our map is excluded, as its gravitational influence becomes negligible due to the $1/r^2$ attenuation and the increasing azimuthal symmetry of the diffuse background at larger distances. For our analysis, we considered pixels within a $4\arcmin$ diameter region with intensities exceeding 18 mJy beam$^{-1}$ (corresponding to a $5\sigma$ detection threshold). Pixels outside this area are masked. The resulting map of gravitational vectors ~--~ displayed at locations where B-field segments are present ~--~ is presented in Figure~\ref{fig:gravity_vector}. Notably, the projected gravitational vectors tend to align toward the filamentary structure, indicating a gravitational pull directed along the filament spine.
Our study aimed to explore the relationship between the gravitational field vector and the B-field orientation and to understand their correlation. To achieve this, we analyzed the angle difference between the two fields, which allowed us to identify areas where the fields were aligned and those where they were not. An offset angle map, by taking the difference between the gravitational field vector and the B-field is shown in the right panel of Figure \ref{fig:gravity_vector}.

\subsection{Filament Properties}
We derived the H$_2$ column density distribution from spectral energy distribution (SED) fitting, as described in Appendix~\ref{appendix b}. The mass in each pixel is calculated as the product of the mean molecular weight per hydrogen molecule ($\mu$), the hydrogen mass (m$\mathrm{H}$), the pixel area ($A$), and the column density. The total filament mass is then obtained by summing over all pixels. Defining the integrated column density as $N(\mathrm{H}2){\mathrm{sum}} = \sum_i N(\mathrm{H}_2)_i$, the total mass can be written as:

\begin{align}
    \hspace{6em} M &= \sum_{i} (\mathrm{\mu} \, \mathrm{m_H} \, \mathrm{A} \, \mathrm{N(H_2)}_i), \\
      &= \mathrm{\mu} \, \mathrm{m_H} \, \mathrm{A} \sum_{i} \mathrm{N(H_2)}_i, \\
      &= \mathrm{\mu} \, \mathrm{m_H} \, \mathrm{A} \, \mathrm{N(H_2)}_{\mathrm{sum}}.
      \label{eq:mass_herschel}
\end{align}

Using this relation, the total filament mass is estimated to be $155.7 \pm 2.9$~M$_{\odot}$. For a filament length of 1.89~pc, the corresponding line mass is $82.0 \pm 1.5$~M$_{\odot}$~pc$^{-1}$.

The critical line mass represents the maximum mass per unit length that can be supported against radial gravitational collapse in an isothermal filament (Ostriker 1964). It is given by:

\begin{align}
    \hspace{6em} \left(\frac{M}{L}\right)_{\mathrm{crit}} = \frac{2 \, c_{\mathrm{s,fil}}^2}{G},
\end{align}

where $c_{\mathrm{s,fil}}$ is the isothermal sound speed, defined as:

\begin{align}
    \hspace{6em} c_{\mathrm{s,fil}} &= \sqrt{\frac{K\mathrm{_B} \, T_{\mathrm{d,fil}}}{\mathrm{\mu} \, m\mathrm{_H}}},
\end{align}

where $T_{\mathrm{d,fil}}$ is the mean dust temperature of $13 \pm 3$~K, obtained as the arithmetic mean over the filament region, and $k_\mathrm{B}$ is the Boltzmann constant. This yields a sound speed of $c_{\mathrm{s,fil}} = 0.20 \pm 0.02$~km~s$^{-1}$.

Using this value, the critical line mass is estimated to be $17.8 \pm 3.5$~M$_{\odot}$~pc$^{-1}$, which is significantly lower than the observed line mass ($82.0 \pm 1.5$~M$_{\odot}$~pc$^{-1}$). This implies that the filament is thermally supercritical, meaning that thermal pressure alone is insufficient to support the filament against its self-gravity. In such a regime, the filament is expected to undergo radial contraction and become gravitationally unstable~\citep{ostriker1964oscillations}.

Physically, when the line mass exceeds the critical value, small perturbations along the filament can grow under gravity, leading to fragmentation into dense cores. These cores can subsequently undergo collapse to form stars. Therefore, the large excess of the observed line mass over the critical value suggests that the filament is highly susceptible to fragmentation.

While the filament is thermally supercritical, the presence of an ordered B-field suggests that magnetic support may play a significant role in regulating its stability. To quantify this, we estimate the mass-to-flux ratio ($\lambda$), which measures the relative importance of gravity and B-fields. Following the formulation of \citet{mouschovias1976stability} and the expression presented by \citet{Fielder_mouschovias_1992}, later commonly adopted in observational studies by \citet{crutcher2004drives}, the mass-to-flux ratio is expressed as,

\begin{align}
\hspace{6em}
    \lambda_\mathrm{obs} = 7.6 \times 10^{-21} 
    \frac{N(\mathrm{H}_2)\,[\mathrm{cm}^{-2}]}{B_\mathrm{3D}\,[\mu\mathrm{G}]},
    \label{eq:m2f}
\end{align}

where $N_\mathrm{H_2}$ is the molecular hydrogen column density and $B_{\mathrm{3D}}$ is the total B-field strength. The total field strength can be estimated as
$B_{\mathrm{3D}} = \frac{4}{\pi} B_\mathrm{POS}$.
Using a mean column density of $1.65 \times 10^{22}$~cm$^{-2}$ and $B_\mathrm{POS} = 153 \pm 6~\mu$G, we obtain $B_{\mathrm{3D}} \approx 195~\mu$G and hence $\lambda \approx 0.64$.

A value of $\lambda < 1$ indicates that the filament is globally magnetically subcritical, implying that B-fields provide significant support against gravitational collapse. Therefore, despite being thermally supercritical, the filament is likely magnetically regulated, with collapse occurring preferentially in localized regions where the critical threshold is exceeded. This behaviour could be facilitated by ambipolar diffusion~\citep{mouschovias1977connection,mouschovias1979ambipolar,mouschovias1991,mouschovias1992ambipolar}. However, we do not observe prominent pinched or hourglass-like B-field morphologies in Cep B. Such morphologies are generally expected in relatively isolated collapsing cores undergoing ambipolar diffusion-driven contraction \citep{Fiedler_1993}. We note that theoretical studies and three-dimensional simulations of ambipolar diffusion-driven fragmentation in filamentary clouds suggest that strong hourglass morphologies may not necessarily develop in clustered or filamentary environments, where fragmentation can be non-uniform \citep{christie2011,mouschovias2026self}.


\subsection{Identification of cores and their properties}
\begin{figure*}
    \centering
    \includegraphics[width=\textwidth]{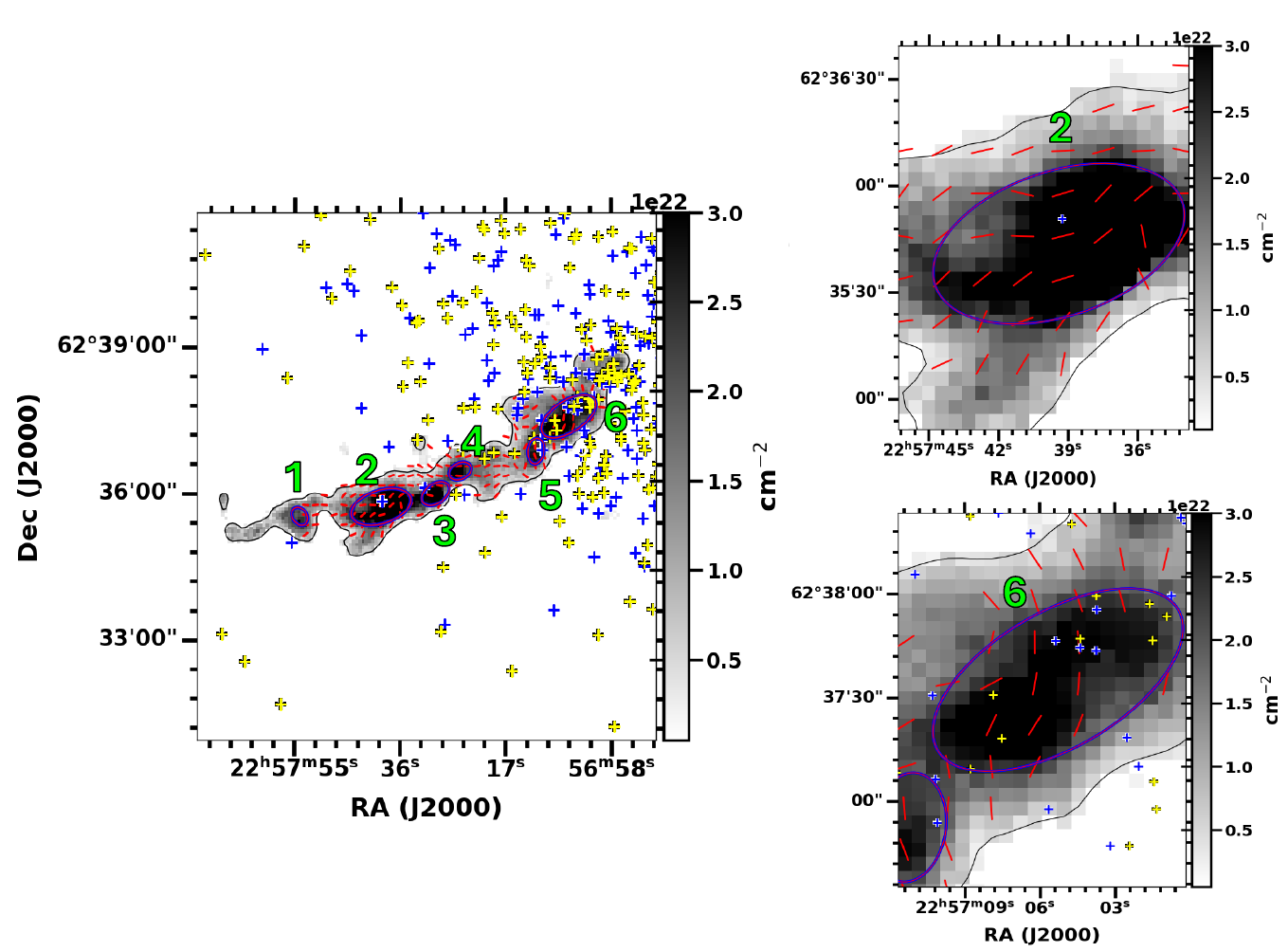}
    \caption{Clumps, indicated by magenta ellipses, were identified in the column density map using the Clumpfind algorithm. The blue `+' sign mark the locations X-ray sources, while the yellow represents the YSOs. The white contour is drawn at 4 $\times$ 10$^{21}$ cm$^{-2}$. Zoomed-in views of cores 2 and 6 are displayed in the right panels.}
    \label{fig:clump_herschel}
\end{figure*}

We identify six cores along the filamentary structure based on their higher column density. To detect these cores, we employed the ClumpFind algorithm from the CUPID package, which is part of the Starlink software. This algorithm was originally developed by \cite{williams1994determining}.
The ClumpFind algorithm outputs five key parameters describing the core properties:
(i) RA and Dec values defining the vertices of the core polygons,
(ii) core centers,
(iii) areas,
(iv) spatial extents in RA and Dec, and
(v) peak dust emission values.
These core boundaries, represented as polygons, are then fitted with ellipses using the SciPy module in Python. The best-fit ellipses provide the semi-major and semi-minor axis dimensions, center coordinates, and position angles (PAs), which are summarized in Table \ref{tab:cores}. The identified cores are visually marked on the Herschel column density map, as shown in Figure \ref{fig:clump_herschel}.
Table \ref{tab:cores} lists additional core properties, including mass, radius, orientation, and inter-core separations projected in 3D space. To obtain the 3D distance, we deproject it by dividing it by a factor of 2/$\pi$ \citep{sanhueza2019alma}. The core masses are derived using Equation \ref{eq:mass_herschel}.

The filament itself is modeled as an isothermal cylinder, where its length corresponds to the height of the cylinder, and its effective radius represents the cylinder's radius. Based on the filament width analysis, we adopt a radius of 0.03 pc and a height of 1.89 pc.

The presence of regularly spaced cores along highly filamentary clouds is broadly consistent with fragmentation driven by the gravitational “sausage” instability of a self-gravitating fluid cylinder. Originally formulated for an incompressible fluid by \cite{chandrasekhar1953problems}, this theory was later extended to isothermal, thermally supported cylinders and to cases including B-fields with various configurations \citep{nagasawa1987gravitational,inutsuka1992self,nakamura1993fragmentation,tomisaka1996collapse}. Unlike spherical Jeans collapse, where perturbations at all wavelengths grow at the same rate, cylindrical collapse favors specific wave numbers that grow fastest, leading to core formation at a characteristic spacing corresponding to the most unstable mode. For an infinite-radius cylinder, this preferred length scale is preserved even in the presence of a B-field aligned with the filament axis \citep{nagasawa1987gravitational}. However, recent theoretical studies and three-dimensional simulations of ambipolar diffusion-driven fragmentation in filamentary clouds suggest that fragmentation may not always be strictly periodic, particularly in clustered or dynamically evolving filamentary environments where B-field morphology can be more complex and core spacings can be non-uniform 
\citep{christie2011,mouschovias2026self}.

The theory, therefore, predicts that filaments should fragment into multiple cores with an approximately periodic spacing set by the wavelength of the fastest growing unstable mode. For an incompressible fluid cylinder, this characteristic wavelength is $\lambda_{\rm max} = 11R$, where $R$ is the cylinder radius \citep{chandrasekhar1953problems}. In contrast, for an infinite isothermal gas cylinder, the fastest-growing mode has a wavelength $\lambda_{\rm max} = 22H$, where the isothermal scale height is $H = c_s(4\pi G \rho_c)^{-1/2}$, with $c_s$ the sound speed and $\rho_c$ the central gas density (R = 0) \citep{nagasawa1987gravitational,inutsuka1992self}. For isothermal cylinders of finite radius embedded in a uniform external medium, the core spacing depends on the ratio $R/H$, approaching $\lambda_{\rm max} = 22H$ for $R \gg H$ and reducing to $\lambda_{\rm max} = 11R$ in the limit $R \ll H$.

For the Cep~B molecular cloud, the central mass density $\rho_c$ was derived assuming a number density of $9.8 \times 10^{4} \mathrm{cm^{-3}}$ (R=0), resulting in $\rho_c \approx 4.6 \times 10^{-16} \mathrm{kg/m^{-3}}$. Using this value, we estimate an isothermal scale height of $H \approx 0.01$ pc. The characteristic filament radius is $R \approx 0.0350$ $\pm$ 0.0005 pc, placing the system in the $R \gg H$ regime. Under these conditions, the theory predicts a characteristic core spacing of $\lambda_{\rm max} = 22H \approx 0.22$ pc, which is in good agreement with the observed core separations reported in Table~\ref{tab:cores}. For a cylinder of infinite radius, this length scale is maintained even in the presence of a B-field parallel to the filament's axis \citep{nagasawa1987gravitational}. 


The theoretically expected fragmentation scale, $\lambda_{\rm max}=22H \approx 0.22$ pc, is comparable to the largest observed core spacings in the filament (0.22 and 0.25 pc), suggesting that gravitational instability sets the characteristic fragmentation scale. However, the remaining spacings are smaller (0.09--0.11 pc), indicating that fragmentation is not perfectly uniform along the filament.

We calculated the theoretical Jeans mass of the observed cores by using the formulation mentioned in \citet{mouschovias2026self},

\begin{align}
    \hspace{6em} M_{\mathrm{J,eff}} = \frac{\pi^{\frac{5}{2}}}{6} \frac{ (\langle C\mathrm{_s} \rangle^2 + \langle C\mathrm{_A} \rangle^2 + \langle \Delta \mathrm{V_{NT}} \rangle^2 )^{3/2}}{ \mathrm{G}^{3/2} \, \mathrm{\rho}^{1/2} } ,
    \label{eq:jeans_mass}
\end{align}

where $C_\mathrm{s}$ is the isothermal sound speed, C$_A$ represents the Alfv{\'e}n velocity, and $\Delta$V$_\mathrm{NT}$ is the non-thermal velocity dispersion for the entire filament. We define an effective Jeans mass by replacing the sound speed with an effective velocity such that $C_{\mathrm{eff}}^2 = C_s^2 + C_A^2 + \Delta V_{\mathrm{NT}}^2$. This $C_\mathrm{s}$ is taken to be 0.20 $\pm$ 0.02 km/s for the entire filament. The C$_A$ and $\Delta$V$_\mathrm{NT}$ are mentioned in Table~\ref{tab:cores}. 


Cores 2 and 6 have masses exceeding $M_{\mathrm{J,eff}}$, indicating that they are gravitationally unstable even after accounting for thermal, turbulent, and magnetic support, and are therefore likely prone to collapse. In contrast, the remaining cores have masses comparable to or below the Jeans mass, suggesting that they may still be supported against gravitational collapse.

\begin{figure}
    \centering
    \includegraphics[width=1\linewidth]{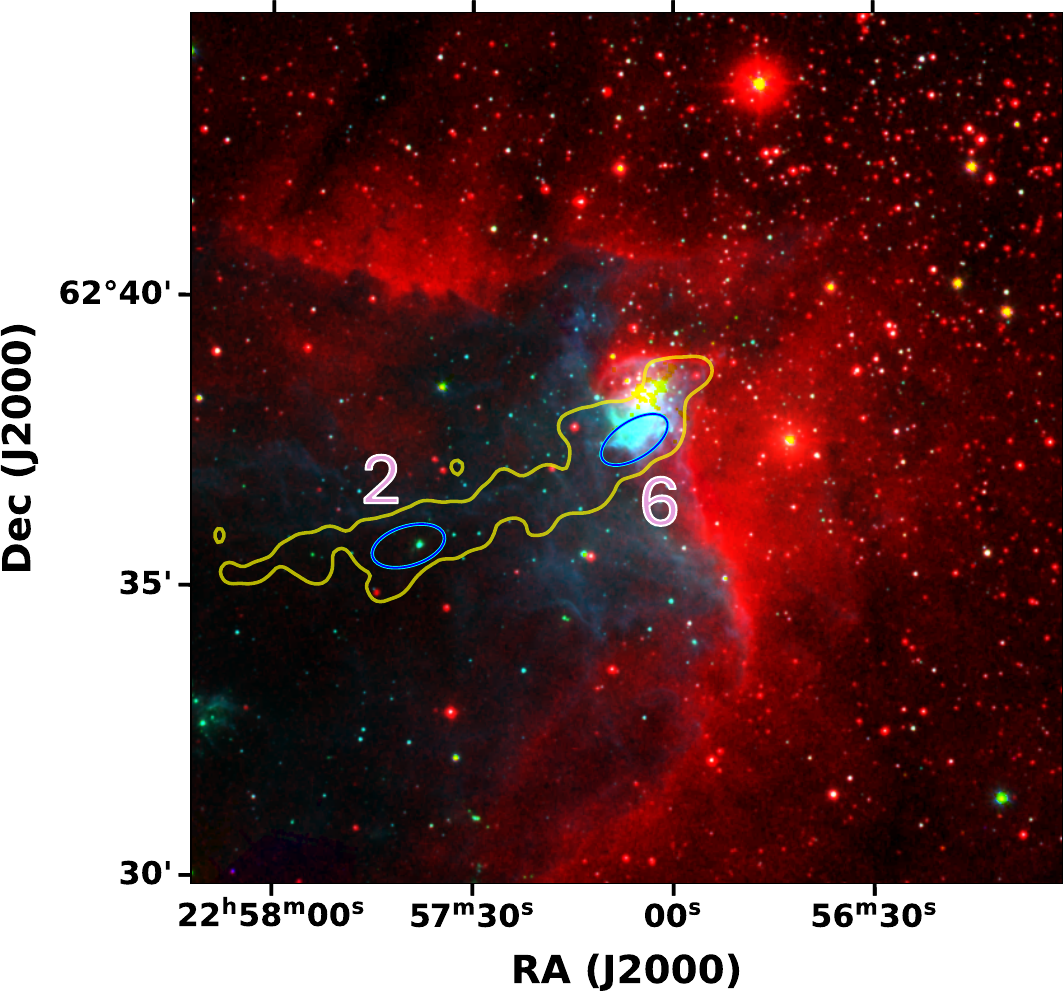}
    \caption{The background image is the three-color composite image of Spitzer 3.6 $\mu$m (blue), 4.5 $\mu$m (green), and DSS2-red (red) data. The yellow contour is the same as Figure~\ref{fig:clump_herschel}, while cores 2 and 6 are marked with blue ellipses.}
    \label{fig:color composite}
\end{figure}

This interpretation is further supported by observational evidence of star formation activity. In particular, Core 6 is associated with X-ray sources and Young Stellar Objects (YSOs) identified by \citet{allen2012spitzer}, which are overplotted on the column density map (Figure~\ref{fig:clump_herschel}). In contrast, Core 2 appears relatively inactive, with only a single X-ray source detected and no clear YSO counterparts. This difference is also reflected in the color composite image (Figure~\ref{fig:color composite}), which shows active clustered star formation toward the head of the cloud, while the tail region, where Core 2 is located, remains comparatively quiescent.

These results suggest that while the Jeans analysis provides a useful indicator of gravitational instability, the evolutionary state of individual cores may vary.

\begin{table*}
    \centering
    \begin{tabular}{ccccccccccc}  
        \toprule
        \textbf{Core} & \makecell{\textbf{RA} \\ \textbf{(deg)}} & \makecell{\textbf{DEC} \\ \textbf{(deg)}} & \makecell{\textbf{Semi-major} \\ \textbf{axis (pc)}} & \makecell{\textbf{Semi-minor} \\ \textbf{axis (pc)}} &
        \makecell{\textbf{M$_\text{Observed}$}\\ \textbf{($M_{\odot}$)}} &
        \makecell{\textbf{M$_\text{Theoretical}$}\\ \textbf{($M_{\odot}$)}} &
        \makecell{\textbf{Core} \\ \textbf{PA (deg)}} &
        \makecell{\textbf{C$_{A}$} \\ \textbf{(m/s)}} &
        \makecell{\textbf{V$_{nth}$} \\ \textbf{(m/s)}}&
        \makecell{\textbf{Intercore Separation} \\ \textbf{(Deprojected) (pc)}} \\
        \midrule
        \multirow{2}{*}{1} & 22:57:38.05 & +62:35:43.83 & 0.041 & 0.031 & 7.0 $\pm$ 3.8 & 112.0 $\pm$ 7.0 & 125.29 & 653$\pm$ 26 & 665$\pm$ 15 & \multirow{2}{*}{\textbf{\bigg\}} 0.25} \\
         & & & & & & & &  \\
        
        \multirow{2}{*}{2} & 22:57:37.44 & +62:35:38.40 & 0.135 & 0.071 & 125.8 $\pm$ 72.4 & 48.0 $\pm$ 16.0 & 16.97 & 398$\pm$ 16 & 722$\pm$ 31 & \multirow{2}{*}{\textbf{\bigg\}}} 0.11 \\
         & & & & & & & &  \\
        
        \multirow{2}{*}{3} & 22:57:29.28 & +62:35:56.40 & 0.067 & 0.042 & 23.3 $\pm$ 14.3 & 50.0 $\pm$ 8.0 & 33.75 & 534$\pm$ 21 & 531$\pm$ 56 & \multirow{2}{*}{\textbf{\bigg\}}} 0.09 \\
         & & & & & & & &  \\
        
        \multirow{2}{*}{4} & 22:57:25.20 & +62:36:28.80 & 0.051 & 0.035 &  10.8 $\pm$ 5.6 & 58.0 $\pm$ 4.0 & 24.05 & 531$\pm$ 21 & 585$\pm$ 18 & \multirow{2}{*}{\textbf{\bigg\}}} 0.22 \\
         & & & & & & & &  \\
        
        \multirow{2}{*}{5} & 22:57:11.28 & +62:36:43.20 & 0.055 & 0.035 & 10.7 $\pm$ 5.4 & 90.0 $\pm$ 17.0 & 82.36 & 540$\pm$ 21 & 750$\pm$ 73 & \multirow{2}{*}{\textbf{\bigg\}}} 0.11 \\
         & & & & & & & &  \\
        
        6 & 22:57:07.20 & +62:37:19.20 & 0.136 & 0.067 & 60.5 $\pm$ 26.7 & 58.0 $\pm$ 17.0 & 32.37 & 439$\pm$ 17 & 729$\pm$ 57 \\  
        \bottomrule
    \end{tabular}
    \caption{Extracted Clump Properties from the FITS Data, Including Mass Estimates.}
    \label{tab:cores}

\end{table*}

\begin{table*}
    
\end{table*}

\subsection{Interaction of internal core pressure with stellar feedback}
In order to shed a light on the impact of stellar feedback on the filament and resultant star formation, and also to infer the impact of B-fields in counteracting the feedback, we estimate the corresponding pressure values. B-field strength and its orientation in the filament with respect to the direction of propagation of I-front can dictate the outcome because the ionized gas can freely stream along the field lines, whereas flows perpendicular to them are hindered by the Lorentz force \citep{tomisaka1992evolution,gaensler1998nature,pavel2012h,van2015shape}. MHD simulations support this picture, showing that B-fields suppress the sweeping up of gas across field lines \citep{krumholz2007magnetohydrodynamic}. 

As time progresses, both gas and magnetic pressures rise at the interacting side (head of the filament facing the ionizing source), increasing the total core pressure until it counteracts the feedback-driven expansion of the I-front. For this to occur, a near pressure equilibrium must be reached, where the internal clump pressure becomes comparable to or exceeds the pressure imparted by feedback from the H{\sc{ii}} region. 
The pressure balance equation is,
\begin{align}
    \hspace{6em} P_\mathrm{B}+P_\mathrm{turb}+P_\mathrm{Tg} = P_\mathrm{Te} + P_\mathrm{rad},
    \label{equlibrium_pressure}
\end{align}
The left-hand side represents the internal pressure of the clump, P$_{\mathrm{clump}}$ = P$_\mathrm{B}$ + P$_{\mathrm{turb}}$ + P$_\mathrm{Tg}$ (Equation (14) of \citep{miao2006triggered}), which is the sum of magnetic pressure ($P_\mathrm{B}$), turbulent (or non-thermal) pressure ($P_{\mathrm{turb}}$), and gas thermal (kinetic) pressure ($P_\mathrm{Tg}$). The latter two components can be combined and expressed as the total molecular gas pressure in the clumps:  
\begin{equation}
\hspace{6em} P_{\mathrm{turb}} + P_\mathrm{Tg} = P_{\mathrm{mol}}.
\end{equation}
The molecular gas pressure ($P_{\mathrm{mol}}$) is estimated following \cite{liu2017alma}:

\begin{equation}
\hspace{6em}P_{\mathrm{mol}} = n_{\mathrm{H_2}} \, k \, T_{\mathrm{eff}},
\end{equation}

\begin{equation}
\hspace{6em}C_{\mathrm{eff}}^{2} = \frac{k \, T_{\mathrm{eff}}}{\mu m_{\mathrm{H}}},
\end{equation}

where $C_{\mathrm{eff}}$ is the effective sound speed, $T_{\mathrm{eff}}$ is the effective temperature, $\mu$ is the mean molecular weight, and $m_{\mathrm{H}}$ is the mass of a hydrogen atom. The effective sound speed is given by

\begin{equation}
\hspace{6em}C_{\mathrm{eff}} = \sqrt{C_s^2 + \sigma_{\mathrm{NT}}^2},
\end{equation}

where $C_s$ is the thermal sound speed and $\sigma_{\mathrm{NT}}$ is the non-thermal velocity dispersion.

For core 6, $C_s = 200 \pm 20~\mathrm{m\,s^{-1}}$ and $\sigma_{\mathrm{NT}} = 729 \pm 57~\mathrm{m\,s^{-1}}$, yielding 
$C_{\mathrm{eff}} = 758.4 \pm 54.8~\mathrm{m\,s^{-1}}$. Using this, the effective temperature is estimated to be 
$T_{\mathrm{eff}} = 194.8 \pm 28.1~\mathrm{K}$. Consequently, the molecular gas pressure is coming out to be P$_{\mathrm{mol}}$ = (29.0 $\pm$ 4.2) $\times$ 10$^{-10}~\mathrm{dyn\,cm^{-2}}$.

We estimated the magnetic pressure using $P_{B} = \frac{B^{2}}{8 \pi}$, with $B = 153 \pm 6 \, \mu$G, which gives $P_{B} = (8.4 \pm 2.9) \times 10^{-10} \, \mathrm{dyn \, cm^{-2}}$. We note that magnetic pressure ($P_\mathrm{B}$ $\propto B^2$), an ordered B-field provides anisotropic support through magnetic tension, in contrast to a tangled field that behaves approximately as an isotropic pressure component.

The right-hand side of Equation~\ref{equlibrium_pressure} corresponds to the feedback pressure,  

\begin{equation}
\hspace{6em}P_{\mathrm{fb}} = P_{\mathrm{Te}} + P_{\mathrm{rad}},
\end{equation}

which is the sum of the pressure from the thermally ionized medium (electron temperature; $P_{\mathrm{Te}}$) and the radiation pressure ($P_{\mathrm{rad}}$), discussed in section \ref{rad_thermal_pressure}.
P$_\mathrm{core6}$ and P$_\mathrm{fb}$ are estimated as  $(37.4 \pm 5.1) \times 10^{-10} \, \mathrm{dyn \, cm^{-2}}$ and $22.6 \times 10^{-10} \, \mathrm{dyn \, cm^{-2}}$. These parameters, shows that P$_\mathrm{core6}$ > P$_\mathrm{fb}$. The core pressure is predominant and be able to stops the expansion of I-front. 


However, since the external pressure in the form of feedback act in favor of gravity, the sum of gravitational and external feedback overwhelms the gas internal pressure, leading to the star formation in core 6. Therefore, we are seeing more active star formation in core 6, compared to others.
\section{Discussion} \label{discussion}
\subsection{B-field Morphology and Radiative Feedback}
The polarization observations of Cepheus B reveal a distinct B-field morphology characterized by a curved, bow-like structure near the irradiated head of the cloud and a more ordered, elongated alignment along the filamentary tail. This transition in field geometry is spatially correlated with the cloud structure, suggesting a direct connection between the B-field configuration and the physical processes shaping the region.


Moving along the filament, the B-field becomes increasingly aligned with the long axis of the structure, as confirmed by the HRO analysis presented in Section~\ref{HRO}. This parallel alignment indicates a strong coupling between the gas and B-field, where the field lines are stretched along the direction of the filament. The transition from a curved morphology at the head to a linear configuration in the tail suggests a systematic reorganization of the B-field across the cloud.

The observed field morphology is consistent with a scenario in which external feedback processes, such as ionization-driven compression, influence both the gas dynamics and the B-field configuration. The bow-shaped structure at the head reflects the impact of radiation on the dense gas, while the aligned field in the tail traces the subsequent redistribution of material along the filament. Similar observational signatures have been reported in other irradiated star-forming regions, including RCW 57A, M16, and IC 1396, where B-fields are observed to bend around ionized boundaries and align with elongated structures~\citep{eswar2017,pattle2018first,soam2018magnetic}.


To further assess the role of stellar feedback and B-fields in shaping the evolution of the filament, we compare the internal pressure of the cores with the external feedback pressure from the adjacent H{\sc ii} region. Our analysis shows that the total internal pressure of core 6, dominated by magnetic and turbulent contributions, exceeds the feedback pressure ($P_{\rm core} > P_{\rm fb}$). This suggests that the core is sufficiently pressurized to resist the expansion of the ionization front, thereby maintaining its structural integrity. 
However, despite this support, the combined effect of self-gravity and external compression can still drive collapse, indicating that feedback may act to enhance, rather than suppress, star formation in localized regions. This highlights a scenario in which B-fields both protect dense structures from erosion and channel feedback-driven flows, while simultaneously allowing gravitational collapse to proceed in the core.

\subsection{Global Stability and Fragmentation of the Filament}

The Cepheus B filament exhibits clear signatures of gravitational instability and fragmentation, as evidenced by its thermally supercritical nature and the presence of multiple dense cores distributed along its length. The observed line mass ($82.0 \pm 1.5$~M$_{\odot}$~pc$^{-1}$) significantly exceeds the critical value expected for an isothermal filament, indicating that thermal pressure alone is insufficient to support the structure against self-gravity. In such a regime, the filament is expected to undergo radial contraction and develop longitudinal density perturbations that grow and lead to fragmentation.

Despite this, the estimated mass-to-flux ratio ($\lambda \approx 0.64$) suggests that the filament remains magnetically subcritical on global scales, implying that B-fields provide substantial support against gravitational collapse. This apparent tension between thermal supercriticality and magnetic subcriticality can be reconciled within the framework of magnetically regulated star formation, where B-fields inhibit global collapse while allowing localized regions to become supercritical through local mass accumulation. As neutrals gradually drift relative to ions, mass accumulates in localized regions, 
eventually producing magnetically supercritical cores embedded within a subcritical envelope. As a result, the filament can maintain overall structural support while still forming stars in localized regions. In this case, although ambipolar diffusion may in principle contribute to fragmentation~\citep{mouschovias1977connection,mouschovias1979ambipolar,mouschovias1991,mouschovias1992ambipolar}, our present observations do not provide direct evidence to conclusively test this scenario. In particular, we do not observe prominent pinched or hourglass-like B-field morphologies near the cores. We note, however, that the hourglass morphology predicted in earlier ambipolar diffusion studies primarily corresponds to relatively isolated fragments or cores \citep{Fiedler_1993}. More recent three-dimensional simulations of ambipolar diffusion-driven fragmentation in filamentary clouds suggest that filamentary structures can fragment non-uniformly and may not necessarily exhibit strong hourglass-like B-field morphologies in clustered or filamentary environments \citep{christie2011, mouschovias2026self}. In addition, projection effects, limited angular resolution, and the dynamically evolving nature of Cep B, including the influence of stellar feedback, may further complicate the observed B-field morphology. A full treatment of non-ideal MHD effects is beyond the scope of the present work.


This interpretation is further supported by the observed core properties in Cep B. A comparison between the observed core masses and the effective Jeans mass, which includes thermal, turbulent, and magnetic support, shows that several cores (Cores 2 and 6) have masses exceeding the critical Jeans threshold. This indicates that these cores are gravitationally unstable and are likely undergoing collapse, which is further supported by the presence of YSOs and X-ray sources. In contrast, the remaining cores have masses comparable to or below the Jeans mass, suggesting that they may still be supported against collapse. 

The spatial distribution of dense cores along the filament provides further insight into the fragmentation process. The quasi-periodic spacing of the identified cores is broadly consistent with the gravitational “sausage” instability of a self-gravitating cylinder. 
Using the derived central density and sound speed, we estimate $\lambda_{\rm max} \approx 0.22$~pc. This value is comparable to the largest observed core separations ($\sim$0.22--0.25~pc), suggesting that gravitational instability sets the characteristic fragmentation scale. However, the observed spacings also show significant variations (0.09--0.25~pc), indicating that fragmentation is not strictly uniform along the filament, consistent with theoretical predictions of ambipolar diffusion-driven fragmentation in filamentary clouds \citep{mouschovias2026self}. Such deviations are expected in realistic molecular clouds, where local variations in density, temperature, B-field strength and orientation, and external feedback can modify the idealized instability pattern. The smaller separations may additionally reflect secondary or hierarchical fragmentation within initially larger unstable segments.

\section{Summary and Conclusion} \label{conclusion}
In this paper, we have presented the dust polarization observations and the $^{13}$CO (J=3-2) spectral line observations using SCUBA-2/POL-2 and HARP on the James Clark Maxwell Telescope (JCMT) of Cepheus B molecular cloud. The main results of our analysis are as follows$\colon$

\begin{itemize}
    \item The JCMT/POL-2 dust continuum map at 850 $\mu$m reveals a prominent, elongated filamentary structure in Cep B, oriented in the Northwest–Southeast (NW–SE) direction. The B-field exhibits a distinct morphology: it curves into a bow-like shape near the head of the filament, while toward the tail, it becomes aligned along the filament spine.
    \item To understand dust grain alignment, the paper analyzes the debiased polarization fraction as a function of total intensity at 850 $\mu$m, finding a power-law exponent of -0.64 $\pm$ 0.02, indicating that dust grains are aligned with the B-field within the dense cloud.
    \item The Structure Function (SF) analysis of the polarization data for the filament yields a ratio of turbulent to ordered B-field ($\langle\Delta B^2\rangle^{1/2}/B_0$) of 0.33 $\pm$ 0.01 and an estimated B-field strength of 181 $\pm$ 9 $\mu$G.
    \item Further analysis using the Autocorrelation Function (ACF) resulted in a turbulent to ordered B-field ratio of 0.47 $\pm$ 0.02 and an estimated B-field strength of 153 $\pm$ 6 $\mu$G in the filament.
    \item The projected gravitational field vectors were analyzed, showing that they are directed toward the filament skeleton, indicating that gravity plays a role in accumulating material from low-density outskirts of the filament to the spine.
    \item The thermal pressure exerted by the ionized gas is approximately $21.4 \times 10^{-10}$ dyns cm$^{-2}$, while the radiation pressure contributions from the O7 star and the B1 star are about $1.2 \times 10^{-10}$ dyns cm$^{-2}$ and $0.34 \times 10^{-12}$ dyns cm$^{-2}$, respectively, in the Cep B H{\sc ii} region.
    \item The HRO analysis reveals a clear density-dependent transition in the relative orientation between the B-field and the filamentary structures in the Cep B filament. At low column densities, the B-field tends to align parallel to the filament, while at higher densities, it shifts to a predominantly perpendicular orientation. 
    \item From the H$_2$ column density distribution, we derive a line mass of 82.0 $\pm$ 1.5 M$_{\odot}$ pc$^{-1}$, significantly exceeding the critical value of 17.8 $\pm$ 3.5 M$_{\odot}$ pc$^{-1}$. This suggests that the filament is gravitationally unstable to radial contraction, providing conditions favorable for subsequent fragmentation.
    \item The observed core spacing spans a range of values, with the largest separations comparable to the theoretically expected fragmentation scale, suggesting that gravitational instability sets the primary fragmentation scale.

    \item The presence of smaller core separations indicates deviations from the idealized model, likely arising from local variations in density, B-field properties, and external feedback, and may point toward hierarchical fragmentation.

    \item The estimated mass-to-flux ratio suggests that the filament is magnetically subcritical on global scales, indicating that B-fields provide significant support against collapse.

    \item Despite global magnetic support, the presence of dense cores and ongoing star formation implies that collapse proceeds locally.
    \item A comparison between the internal core pressure and external feedback pressure shows that the total internal pressure of Core~6, dominated by magnetic and turbulent contributions, exceeds the feedback pressure ($P_{\rm core} > P_{\rm fb}$). This indicates that the core can resist direct compression from the ionization front and maintain its structural integrity.
    \item Overall, the Cepheus B filament represents a system in which gravity drives fragmentation, B-fields regulate its evolution, and external feedback influences both morphology and star formation.
\end{itemize}

\section{Acknowledgements}
SP thanks the DST-INSPIRE Fellowship (No. IF200294) from the Department of Science and
Technology (DST) for supporting the Ph.D. program. CE acknowledges the support from
a Core Research Grant (CRG; sanction order number CRG/ 2023/008710) awarded by the
Anusandhan National Research Foundation (ANRF) under Science and Engineering Research
Board (SERB), Govt. of India. CE acknowledges the financial support from grant RJF/2020/
000071 as a part of the Ramanujan Fellowship awarded by the Science and Engineering Research Board (SERB), Department of Science and Technology (DST), Govt. of India. Y. M. acknowledges the support of NSFC grant 12303033. MRS acknowledges the support from the Department of Space, Government of India. PS was partially supported by a Grant-in-Aid for Scientific Research (KAKENHI Number JP23H01221) of JSPS. The James Clerk Maxwell Telescope (JCMT) is operated by the East Asian Observatory on behalf of the National Astronomical Observatory of Japan, the Academia Sinica Institute of Astronomy and Astrophysics, the Korea Astronomy and Space Science Institute, the National Astronomical Research Institute of Thailand, and the Center for Astronomical Mega-Science, with additional support from the National Key R\&D Program of China (Grant No. 2017YFA0402700). The Science and Technology Facilities Council of the United Kingdom, along with participating universities and organizations in the United Kingdom and Canada, also provide funding support. The construction of SCUBA-2 was additionally supported by the Canada Foundation for Innovation. The authors respectfully acknowledge the cultural significance and enduring reverence of Maunakea within the indigenous Hawaiian community, and express gratitude for the privilege of conducting observations from this mountain.
\section{Data Availability}
All data supporting the conclusions of this paper are available from the corresponding and first author upon reasonable request.
\appendix

\bsp	
\label{lastpage}\
\bibliographystyle{mnras}
\bibliography{mnras_template}
\bsp
\section{Herschel Column density and Temperature map}\label{appendix b}
Foreground extinction can lead to a bias in the measured column density. This occurs because foreground extinction can cause the light from a background source to be absorbed, reducing the amount of light that reaches the observer. This reduction in light can lead to an overestimation of the column density, which can bias the measurement. This suggests that foreground dust can significantly impact the measured column density, especially in diffuse areas, and it's important to consider and minimize the effects of foreground contamination when estimating the temperature and column density from continuum data. We have tried to estimate the temperature and column density from the Herschel 250, 350, and 500 $\mu$m continuum data, but we
found that the measured column density is generally too high ($\approx$ 10$^{22}$ cm$^{-2}$) in the diffuse areas. The JCMT data, with its filtering of large-scale extended emission, appears to have reduced levels of foreground contamination. Therefore, we are determining the column density of the target using the JCMT 850 $\mu$m continuum data.
The column density and Temperature maps are produced using the following relation \cite{kauffmann2008mambo}
\begin{align}
    s_{\nu^{\text{beam}}} = & \left(\frac{N_{\mathrm{H_2}}}{2.02 \times 10^{20}~\mathrm{cm}^{-2}}\right) 
    \times \left(\frac{K_{\nu}}{0.01~\mathrm{cm}^2~\mathrm{g}^{-1}}\right) 
    \times \left(\frac{\theta_{\mathrm{HPBW}}}{10''}\right)^2 \notag \\
    & \times \left(\frac{\lambda}{\mathrm{mm}}\right)^{-3} 
    \times \left(e^{1.439 \left(\frac{\lambda}{\mathrm{mm}}\right)^{-1} \left(\frac{T}{10~\mathrm{K}}\right)^{-1}} -1\right)
\end{align}

s$_\nu^{\text{beam}}$ denotes the flux density per beam, $N{\mathrm{H}2}$ is the column density, and $K\nu = 0.1 \left(\frac{\nu~(\mathrm{GHz})}{1000}\right)^\beta$ represents the dust opacity coefficient, where $\nu$ is the frequency in GHz and $\beta$ is the dust opacity index, assumed to be 2. The parameter $\theta_{\mathrm{HPBW}}$ (arcsec) denotes the half-power beam width of the instrument, with values for different bands as follows: PACS 160~$\mu$m (11.4$\arcsec$), SPIRE 250~$\mu$m (18.1$\arcsec$), SPIRE 350~$\mu$m (24.9$\arcsec$), SPIRE 500~$\mu$m (36.4$\arcsec$), and JCMT 850~$\mu$m (4$\arcsec$). Spectral Energy Distribution (SED) fitting was performed across these bands, yielding the column density and dust temperature as best-fit parameters. PACS 160 $\mu$m intensity values are in Jy/beam, and SPIRE 250 $\mu$m, SPIRE 350 $\mu$m, SPIRE 500 $\mu$m intensity values are in Jy/beam and the JCMT 850 $\mu$m is in mJy/beam. Further, each have different resolution. So we convolved all the datas to same resolution of 36.4$\arcsec$ and converted the units of all the maps to mJy/beam.
so we got the column density and temperature map. But to produce high resolution column density map we need to consider the JCMT 850 $\mu$m data. By utilizing the temperature map with low resolution (36.4$\arcsec$) and the dust emission map with high resolution (4$\arcsec$), we can calculate the high-resolution column density map through the application of equation.
\end{document}